\documentclass[prb,  twocolumn, showpacs, superscriptaddress,longbibliography]{revtex4-1}
\pdfoutput=1
\usepackage{amsmath, amsfonts, amssymb, mathrsfs}
\usepackage{graphicx, color}
\usepackage{xargs}                      
\usepackage[pdftex,dvipsnames]{xcolor}  
\usepackage[caption=false]{subfig}
\usepackage{tikz}
\usepackage{float}
\usepackage{tabularx}
\usepackage{booktabs}
\usepackage{bm}
\usepackage{braket}
\usepackage{epstopdf}
\usepackage{dsfont}
\usepackage{float}
\usepackage{mathtools}
\usepackage{hyperref}
\hypersetup{
   colorlinks,
   citecolor=blue,
   filecolor=blue,
   linkcolor=blue,
   urlcolor=blue,
   breaklinks=true
}

\definecolor{grey}{rgb}{.6,.6,.6}

\newcommand{\Aa}{\mathcal{A}}

\newcommand{\nA}{\overline{\mathcal{A}}}

\begin{document}
\title{Many-body localization in a fragmented Hilbert space}
\author{Lo\"ic Herviou}
\affiliation{Department of Physics, KTH Royal Institute of Technology, Stockholm, 106 91 Sweden}
\affiliation{Institute of Physics, Ecole Polytechnique F\'{e}d\'{e}rale de Lausanne (EPFL), CH-1015  Lausanne, Switzerland}
\author{Jens H. Bardarson}
\affiliation{Department of Physics, KTH Royal Institute of Technology, Stockholm, 106 91 Sweden}
\author{Nicolas Regnault}
\affiliation{Joseph Henry Laboratories and Department of Physics, Princeton University, Princeton, New Jersey 08544, USA}
\affiliation{Laboratoire de Physique de l'\'Ecole normale sup\'erieure, ENS, Universit\'e PSL, CNRS, Sorbonne Universit\'e, Universit\'e Paris-Diderot, Sorbonne Paris Cit\'e, Paris, France}
\begin{abstract}
We study many-body localization (MBL) in a pair-hopping model exhibiting strong fragmentation of the Hilbert space.
We show that several Krylov subspaces have both ergodic statistics in the thermodynamic limit and a dimension that scales much slower than the full Hilbert space, but still exponentially.
Such a property allows us to study the MBL phase transition in systems including more than $50$ spins.
The different Krylov spaces that we consider show clear signatures of a many-body localization transition, both in the Kullback-Leibler divergence of the distribution of their level spacing ratio and their entanglement properties.
But they also present distinct scalings with system size.
Depending on the subspace, the critical disorder strength can be nearly independent of the system size or conversely show an approximately linear increase with the number of spins.
\end{abstract}
\maketitle

\section{Introduction}

Many-body localization (MBL) and its transition\citep{Gornyi2005, Basko2006, Abanin2017, Alet2018} have been the subjects of numerous studies over the recent decades.
They are directly related to core physical concepts and properties of the physics of closed quantum systems, namely thermalization, transport, and the effects of disorder.
Interacting systems at weak disorder thermalize and present ergodic features seemingly following the so-called strong eigenstate thermalization hypothesis (ETH)\citep{Deutsch1991, Srednicki1994, Rigol2008}.
It states that, at high energy, a generic closed quantum system has all its eigenstates display thermal values for all local observables.
At strong disorder, on the other hand, theoretical arguments and numerical studies show a breakdown of the ETH in one dimensional systems, arising from emergent integrability and approximate integrals of motions\citep{Serbyn2013, Huse2014, Chandran2015, Ros2015, Bera2015, Rademaker2016, Imbrie2016, Imbrie2016-2, OBrien2016}.
The high-energy eigenstates are then characterized by low entanglement, following an area law instead of a volume law as in the ergodic phase\citep{Znidaric2008, Bardarson2012, Bauer2013, Kjall2014, Luitz:2016hr,Yu:2016gb}.
While the existence of the MBL phase is now generally well accepted, despite a recent debate\citep{Prosen2019, Abanin2019, Panda2020, Sierant2020, Sierant2020-2, Sirker2020, Sirker2020-2}, describing the transition from the ergodic phase to the localized phase remains a numerical and theoretical challenge.
Indeed, the physics of MBL arises from the rich interplay of various phenomena: many-body interactions allowing for non-integrability, (strong) disorder leading to localization, and high-energy physics. 
Many theoretical studies rely on phenomenological renormalization group arguments, based on various physical arguments on thermalization and predictions of random matrix theory\citep{Potter2015, Vosk2015, Zhang2016, Monthus2017, Dumitrescu2017, Thiery2017, Thiery-2017-long, Goremykina2018, Igloi2018, Dumitrescu2018}.
Numerical studies are also limited by the complexity of the problems: matrix-product-states based approaches\citep{Khemani2016, Yu2016, Lim2016, Devakul2016, Serbyn2016, Doggen2018} perform well deep inside the MBL phase but become unreliable close to the transition due to rapidly increasing entanglement, leaving exact diagonalization and variants thereof, with its generally limited system sizes, as the main source of exact numerical resources\citep{Berkelbach2010, Pal2010, Luitz2015, Pietracaprina2017, Lezama2017, Pietracaprina2018, Sierant2020-2, Herviou2019}.
The existence of MBL, as a means to break the strong ETH beyond integrability, spurred the growth of interest in other phenomenas leading to such a breakdown.\citep{DeRoeck2014, Schiulaz2015, Yao2016, Smith2017, Brenes2018} 
Two majors archetypes have emerged: many-body quantum scars\citep{Bernien2017, Modgalya2018, Schecter2018, Turner2018, Turner2018-2, Choi2019, Ho2019, Moudgalya2020-eta, Mondragon2020} and Krylov fragmentation\citep{Znidaric2013, Moudgalya2019, Moudgalya2019-2, Iadecola2019, Khemani2020, Sala2020, Yang2020}.
Systems with quantum scars present a set of measure zero of highly excited non-thermal eigenstates, typically characterized by a sub-volume law entropy.
The other eigenstates remain thermal.
The presence of these states has especially strong consequences on nonequilibrium dynamics in such systems, with partially suppressed thermalization\citep{Bernien2017, Turner2018}. 
Depending on the initial state, time-evolution under a Hamiltonian presenting these scar states can typically present much slowler relaxation of observables towards the thermal equilibrium states, with slowly suppressed revivals at long times.
Generic methods to embed such states into a thermal spectrum have been proposed\citep{Bull2019, Bull2020} and scars have been proved to be resilient to the effect of disorder\citep{Mondragon2020}. 
More relevant to this work is the concept of Krylov subspaces or Hilbert space fragmentation\citep{Znidaric2013, Moudgalya2019, Moudgalya2019-2, Iadecola2019, Khemani2020, Sala2020, Yang2020, Scherg2020}.
Due to the interplay between different $U(1)$ symmetries such as charge and dipole conservation, each symmetry sector of the Hilbert space shatters into an exponential number of sectors or \emph{Krylov subspaces} that are not connected by the Hamiltonian dynamics.
Importantly, these subspaces are not fully labeled by quantum numbers. 
The exponential number of small disconnected sectors leads to anomalous and effectively localized dynamics\citep{Iadecola2019, Khemani2020, Scherg2020}.
Conversely, exponentially large Krylov subspaces have recently been the subject of several studies.\citep{Moudgalya2019, Yang2020}
Remarkably, in the same model and in the absence of disorder, some of these Krylov subspaces follow the ETH, while other subspaces have completely integrable statistics.

A natural question for these systems is the effect of disorder on these Krylov subspaces, and in particular whether it can preserve the fragmented nature of the Hilbert space, and lead to a localization of the different ergodic subspaces.
More importantly, we identify sets of ergodic subspaces whose dimension grows much slower than the total Hilbert space dimension, albeit still in an exponentional fashion.
This gives us the possibility to investigate through exact (and full) diagonalization one-dimensional systems of unprecedented physical sizes.
A similar approach was proposed in Ref.~\onlinecite{Pietracaprina2019}: in the conventional $XXZ$ model, strong disorder permits to approximately separate the Hilbert space into quasi-independent subspaces.
There, the separation is \emph{only} a strong-disorder induced approximation.
In our model, it is \emph{exact} at all disorder strengths.
Our approach also shares some similarities with the studies of frustrated models such as quantum dimers whose Hilbert space shows slow exponential scalings and signs of MBL even in two dimensions\citep{Theveniaut2020, Pietracaprina2020}.\\

The outline of our paper is as follows.
In Section \ref{sec:model}, we introduce the pair-hopping model and its main properties and symmetries.
Section \ref{sec:Krylovsubspaces} is dedicated to Krylov subspaces.
After a formal definition and a discussion of the Krylov subspaces studied in Ref.~\onlinecite{Moudgalya2019}, we introduce our slowly-growing ergodic subspaces.
We study the level spacing ratio statistics in the pair-hopping model in Section \ref{sec:LS}.
We discuss the distribution of the level spacing ratio of the full Hilbert space, showcasing the need to consider the individual Krylov subspaces.
To probe the ETH-MBL transition within the Krylov subspaces, we rely on the Kullback-Leibler divergence of the level ratio distribution and the reference GOE and Poisson distributions\citep{Wigner1955, Pal2010, Alet2018}.
We identify the critical disorder strength as the point of maximum confusion.
We show that our Krylov subspaces present all signs of the MBL phase transitions. 
The different critical disorders nonetheless have radically different scalings with system size, ranging from quasi absent finite-size effects to an approximately linear shift (within the size we have access to).
Our results underline the importance of the structure of the Hilbert space in the behavior of the MBL phase transition.
We also briefly discuss the existence of a mobility edge\citep{Gornyi2005, Basko2006, Luitz2015}.
We then turn to the von Neumann entanglement entropy (vNEE) of highly excited states in Section \ref{sec:ententropy}.
The critical disorder strengths, identified by the maximum of the standard deviation of the mid-chain entanglement entropy\citep{Kjall2014}, are in qualitative agreement with our previous results.
We also carefully discuss the scaling of the vNEE with subsystem size which present unusual plateaus due to the strongly constrained nature of our model.

\section{Pair hopping model in a transverse field}\label{sec:model}
The pair hopping model\citep{Seidel2003, Moudgalya2019, Moudgalya2019-2} is an interacting model of spinless fermions on a one-dimensional lattice whose Hamiltonian is given by:
\begin{equation}
\tilde{H}_\mathrm{PP} = \sum\limits_j J_j (c^\dagger_j c_{j+1} c_{j+2} c^\dagger_{j+3} + h.c.), \label{eq:pairHoppingHamiltonian}
\end{equation}
where $c_j$ ($c_j^{\dagger}$) is the fermionic annihilation (creation) operator on site $j$, $J_j$ are site-dependent pair hopping terms and we fix the number of sites to $L$.
We take $J_j$ uniformly sampled in $[0.9, 1.1]$ in order to break inversion symmetry and translation symmetry. 
For convenience, we perform a Jordan-Wigner transform, and work with the spin-$\frac{1}{2}$ Hamiltonian
\begin{equation}
H_\mathrm{PP} = \sum\limits_j J_j (\sigma^+_j \sigma^-_{j+1} \sigma^-_{j+2} \sigma^+_{j+3} + h.c.),\label{eq:pairHoppingHamiltonianSpin}
\end{equation}
and consider either open boundary conditions (OBC) or periodic boundary conditions (PBC) in the spin basis.
PBC in the spin basis are not equivalent to fermionic PBC due to the presence of the fermionic string, but the observables we consider in the remainder of this article are largely unaffected.
This model preserves the total polarization (or charge in the fermionic language)  
\begin{equation}
P_\mathrm{tot}=\sum\limits_{j=1}^L \sigma^z_j \label{eq:P-sym}
\end{equation}
and the dipole moment (or center-of-mass position) defined as
\begin{equation}
 C=\left\lbrace \begin{tabular}{cc}
 $\sum\limits_j j \sigma^z_j$ & if OBC\\
 $\exp \frac{i \pi}{L} \sum\limits_j j \sigma^z_j$ & if PBC.\\
 \end{tabular}\right.\label{eq:C-sym}
\end{equation} 
Additionally, for $L$ even with PBC and all $L$'s with OBC, the pair hopping terms preserve the sublattice symmetry, i.e., $P_\mathrm{o}- P_\mathrm{e}$ where $P_\mathrm{e}$ ($P_\mathrm{o}$) is the total charge of the even (odd) sites, and therefore these two charges are also conserved quantities.
We denote with $p_\mathrm{e}$, $p_\mathrm{o}$, $p_\mathrm{tot}$ and $c$ the quantum numbers respectively associated to $P_\mathrm{e}$, $P_\mathrm{o}$, $P_\mathrm{tot}$ and $C$.
Similar pair-hopping terms appear naturally in different experimentally relevant set-ups, such as electrons in a Landau level\citep{Bergholtz2006, Bergholtz2008, Moudgalya2019-2} and in the Wannier-Stark problem\citep{Wannier1962, 	vanNieuwenburg9269, Schulz2019}.\\

We introduce disorder in the form of a transverse field (a random on-site chemical potential in the fermionic language), resulting in the total hamiltonian 
\begin{equation}
H = H_\mathrm{PP} + \sum\limits_{j=1}^L W_j \sigma^z_j,
\end{equation}
where $W_j$ is taken uniformly in $[-W, W]$.
This disorder does not break any of the identified symmetries.\\

To better understand the physics of the hopping terms, it is convenient to represent the system in terms of pair of spins, using the notations introduced in Ref.~\onlinecite{Moudgalya2019}.
For convenience, we assume $L$ even in the rest of the paper.
Let the local Hilbert space be defined by
\begin{equation}
\sigma^z \Ket{1} = \Ket{1},\ \sigma^z \Ket{0} = - \Ket{0}.
\end{equation}
The system is composed of $N=L/2$ pairs of spins, decomposed in the basis
\begin{equation}
\Ket{\uparrow} = \Ket{10},\ \Ket{\downarrow} = \Ket{01},\ \Ket{-}=\Ket{00}\ \text{and } \Ket{+}=\Ket{11}.
\end{equation}
We denote $\Ket{\uparrow}$ and $\Ket{\downarrow}$ as pseudo-spins, $\Ket{+}$ and $\Ket{-}$ as fractons and the combination $\Ket{+-}$ and $\Ket{-+}$ as dipoles.
While $\sigma^z$'s are still diagonal in this basis, the hopping terms of Eq.~\eqref{eq:pairHoppingHamiltonian} lead to some complex algebra.
The transformation rules are the following:
\begin{equation}
\Ket{\uparrow \downarrow} \leftrightarrow \Ket{\downarrow \uparrow } \label{eq:pseudospinExchange}
\end{equation}
\begin{equation}
\Ket{\uparrow + -} \leftrightarrow \Ket{+ - \uparrow }, \quad \Ket{\downarrow - +} \leftrightarrow \Ket{- + \downarrow} \label{eq:dipoleMovement}
\end{equation}
\begin{equation}
\Ket{+-+} \leftrightarrow \ket{\uparrow + \downarrow}, \quad \Ket{-+-} \leftrightarrow \ket{\downarrow - \uparrow}\label{eq:fractonSource}
\end{equation}
Pseudo-spins exchange with each other (Eq.~\eqref{eq:pseudospinExchange}), and dipoles can move each in one flavor of pseudo-spins (Eq.~\eqref{eq:dipoleMovement}).
Conversely, well-chosen trio of fractons can transform into a fracton and a pair of pseudo-spins, and reversely the presence of a fracton may lead a pair of pseudo-spins to transform into a pair of fractons.

\section{Krylov subspaces}\label{sec:Krylovsubspaces}
In this Section, we introduce the notion of Krylov subspaces, vector subspaces of the symmetry-resolved Hilbert space which are stable under the application of the Hamiltonian.
In particular, we present several families of Krylov subspaces, ergodic at zero disorder despite their simple structure, whose dimension grows slowly with system size.

\subsection{Definition}
A natural approach to find the stable subspaces induced by the fragmentation of the Hilbert space is to work with Krylov subspaces\citep{Znidaric2013, Moudgalya2019, Iadecola2019, Khemani2020, Sala2020, Yang2020}. 
Generally, the Krylov space $\mathcal{K}_1(\Ket{\Psi}, H)$ generated by the state $\Ket{\Psi}$ and the Hamiltonian $H$ is defined as follows:
\begin{equation}
\mathcal{K}_1(\Ket{\Psi}, H) = \text{Span} (\Ket{\Psi}, H\Ket{\Psi}, H^2 \Ket{\Psi} ...).
\end{equation}
This definition has the drawbacks of being numerically unstable for generic Hamiltonians and requiring the orthogonalization of a set of large vectors.
Working in the configuration basis
\begin{align}
\mathcal{B} &= \{ \bigotimes\limits_{j=1}^N \Ket{\alpha_j} \text{ with } \alpha_j \in \{\uparrow,\, \downarrow,\, +,\, -\}  \}\\
&= \{ \bigotimes\limits_{j=1}^L \Ket{\alpha_j} \text{ with } \alpha_j \in \{0,\, 1\}  \},
\end{align}
it is more convenient to use the modified definition:
\begin{equation}
\mathcal{K}_2(\Ket{\Psi}, H) = \text{Span} ( \{ \Ket{w} \in \mathcal{B} \vert \exists n \in \mathbb{N}, \Braket{w \vert H^n \vert \Psi} \neq 0 \}).
\end{equation}
For simplicity, we consider in the following $\Ket{\Psi} \in \mathcal{B}$, i.e., a product state in the $\sigma^z$'s basis.
Then, seeing the Hamiltonian as a graph in the configuration space $\mathcal{B}$, where each node is an element of the basis and each link a non-zero coefficient of the Hamiltonian, $\mathcal{K}_2(\Ket{\Psi}, H) $ is simply the connected subgraph including $\Ket{\Psi}$.
This definition is numerically stable and straightforward to compute.
Note that contrary to $\mathcal{K}_1$, $\mathcal{K}_2$ strongly depends on the basis $\mathcal{B}$, which should be specifically crafted for the studied model.
Our choice for $\mathcal{B}$ for the pair-hopping model ensures the following property:
\begin{equation}
\mathcal{K}_2(\Ket{\Psi}, H) = \bigcup\limits_{\forall J_j, W_j} \mathcal{K}_1(\Ket{\Psi}, H) .
\end{equation}
In fact, $\mathcal{K}_1$ and $\mathcal{K}_2$ here nearly always coincide\footnote{$\mathcal{K}_1$ and $\mathcal{K}_2$ here will differ only when $J_j=0$ for a site $j$ in the chain. This forms a set of measure $0$ in the space of Hamiltonians.}.
We therefore only work with $\mathcal{K}_2(\Ket{\Psi}, H)$ and drop the subscript and explicit dependence on the Hamiltonian in what follows.

\subsection{Ergodic Krylov subspace with a single pair of dipoles}\label{sec:Krylov-basic}

In Refs.~\onlinecite{Moudgalya2019, Yang2020}, the authors observed that exponentially large subspaces with either Poissonian or ergodic statistics might coexist in the absence of disorder.
An especially convenient family of  exponentially large ergodic subspaces were generated by the pair of dipoles $\Ket{-+ +-}$ or $\Ket{+--+}$ in a sea of pseudo spins $\Ket{\uparrow}$ and $\Ket{\downarrow}$.
In the absence of dipoles, the pair-hopping Hamiltonian acts in the sea of pseudo spins, with conservation of each flavour of pseudo-spins. 
Introducing a pair of dipoles breaks down integrability, as each can only move through a single flavour of pseudo-spins.
The pair-hopping Hamiltonian acting on this subspace conserves the number of each flavor of pseudo-spin and fracton, leading to a remarkably simple structure of the corresponding Krylov subspace.

The first Krylov subspace we consider is therefore generated by the state
\begin{align}
\Ket{\Psi_n^1}&=\Ket{(\uparrow \downarrow)^n - ++- (\uparrow \downarrow)^n} \label{eq:KrylovInfty}.
\end{align}
where $(w)^n$ marks that we repeat $n$ times the sequence $w$.
The system is comprised of $N=4n+4$ pairs of spins and $-++-$ is placed exactly at the center of the chain.
This Krylov space was shown to be ergodic in the absence of a transverse field\citep{Moudgalya2019} and its dimension can be readily computed.
For simplicity, we first consider periodic boundary conditions.
The dipoles $\Ket{-+}$ and $\Ket{+-}$ can be seen as separating the pseudo-spins into two sequences: either in between $\Ket{-+}$ and $\Ket{+-}$ or outside of them.
We denote by $N_\uparrow=2n$ the conserved total number of pseudo-spins $\uparrow$, and by $n_\uparrow$ (resp. $n_\downarrow$) the number of $\uparrow$ (resp. of $\downarrow$) between $\Ket{-+}$ and $\Ket{+-}$.
As a concrete example, the state
\begin{equation}
\lvert \uparrow \uparrow \downarrow \uparrow \uparrow - + \overbrace{\uparrow \underbrace{\downarrow \downarrow \downarrow}_{n_\downarrow = 3} \uparrow}^{n_\uparrow + n_\downarrow = 5} + -  \downarrow \downarrow \uparrow \downarrow\uparrow \downarrow \rangle \label{eq:ExampleState}
\end{equation}
belongs to $\mathcal{K}(\Ket{\Psi_{n}^1}, H_\mathrm{PBC})$ with $n=4$, $N_\uparrow=8$, $n_\uparrow = 2$ and $n_\downarrow=3$.
The dimension of the Krylov subspace is simply given by:
\begin{align}
\text{dim }\mathcal{K}(\Ket{\Psi_{n}^1})_{\mathrm{PBC}} &= \sum\limits_{n_\uparrow, n_\downarrow=0}^{N_\uparrow}  \binom{2N_\uparrow - n_\uparrow - n_\downarrow}{N_\uparrow-n_\uparrow} \binom{n_\uparrow + n_\downarrow}{n_\downarrow} \notag \\
&= \frac{(2N_\uparrow+1)!}{N_\uparrow!^2}  \approx \frac{2^{N} \sqrt{N}}{8e \sqrt{2\pi}},\label{eq:dimPsiPBC}
\end{align}
using the Chu-Vandermonde identity for simplification, and the Stirling's approximation for the factorial.
The dimension of this Krylov subspace therefore scales as $2^N=\sqrt{2}^L$, i.e., much slower than the full Hilbert space's dimension, which scales as $2^L$.\\

For OBC, the two dipoles separate the chain in three.
There is a fixed number $N_\uparrow^L=n$ of pseudo-spins $\uparrow$ ($\downarrow$) to the left (right) of the dipoles.
The $n_\downarrow$ $\downarrow$-pseudo-spins in between $\Ket{-+}$ and $\Ket{+-}$ originally came from the left of $\Ket{-++-}$ via acting by $H_\mathrm{OBC}$; similarly the $n_\uparrow$ $\uparrow$ in between pseudo-spins came from the right. 
Hence the state represented in Eq.~\eqref{eq:ExampleState} also belongs to $\mathcal{K}(\Ket{\Psi_{n}^1})_{\mathrm{OBC}}$ with $N_\uparrow^L=4$, $n_\uparrow = 2$ and $n_\downarrow=3$.
The dimension of $\mathcal{K}(\Ket{\Psi_{n}^1})_{\mathrm{OBC}}$ is thus given by
\begin{widetext}
\begin{equation}
\text{dim }\mathcal{K}(\Ket{\Psi_{n}^1})_{\mathrm{OBC}} = \sum\limits_{n_\uparrow, n_\downarrow=0}^{N^L_\uparrow}	  \binom{2N^L_\uparrow-n_\uparrow}{N^L_\uparrow} \binom{2N^L_\uparrow-n_\downarrow}{N_\uparrow^L} \binom{n_\uparrow + n_\downarrow}{n_\downarrow}.\label{eq:dimPsiOBC}
\end{equation}
\end{widetext}
Using twice the Chu-Vandermonde equality (see App.~\ref{app:Krylovinfty}), we obtain
\begin{equation}
\text{dim }\mathcal{K}(\Ket{\Psi_{n}^1})_{\mathrm{OBC}} = \binom{4N_\uparrow^L+1}{2N_\uparrow^L} \approx \frac{2^{N}}{8\sqrt{2 \pi N}}.\label{eq:dimPsiOBC2}
\end{equation}
The exponential scaling is similar, albeit with a more favorable prefactor.\\

For both boundary conditions, the reduced Hilbert space scaling is not due to an extremal choice of quantum numbers (such as the linear scaling of the one particle sector of a particle number conserving $U(1)$ model).
It is a simple consequence of the presence of an extensive number ($\propto N$) of freely exchanging pseudo-spins (composed of two real spins) that make most of the degrees of freedom.
A table summarizing the properties of the Krylov subspace for numerically relevant values of $N$ can be found in App.~\ref{app:Krylovinfty}.
Due to the significantly smaller Krylov space's dimension, we focus on OBC for this family.

\subsection{Slowly-growing ergodic Krylov subspaces}\label{sec:Krylov-slow}
It is possible to construct a series of ergodic subspaces with even more favorable scaling with system size, which cannot be mapped to any simple quasi-particle picture.
Working with sets of $\Ket{\uparrow \downarrow}$ and pairs of dipoles $\Ket{- + + -}$ allows us to keep a simple analytical structure of the Krylov subspace while working in the sector $(p_\mathrm{o}, p_\mathrm{e}, c)=(0, 0, 0)$.
If we fix the number of dipoles to be constant when increasing system size, the dimension of the resulting Krylov space ultimately grows as $2^N$. 
A natural way to go beyond this limit is to alternate between a finite number of pairs of pseudo-spins and the set of two dipoles.
The less pseudo-spins, the slower the growth of the Hilbert space.
The pair-hopping Hamiltonian in Eq.~\eqref{eq:pairHoppingHamiltonian} cancels the state $\Ket{- + + - - + + - ...}$, which therefore forms a Krylov subspace of dimension $1$.
We define the state $\Ket{\Phi_n^m}$ for periodic boundary conditions as:
\begin{equation}
\Ket{\Phi_{n}^m} = \Ket{ \left((\uparrow \downarrow)^n - + + - \right)^m },\label{eq:Secondfamily}
\end{equation}
where $(w)^m$ again marks that we repeat $m$ times the sequence $w$.
Hence,
\begin{equation}
\Ket{\Phi_{3}^2} = \Ket{\uparrow \downarrow \uparrow \downarrow \uparrow \downarrow - + + - \uparrow \downarrow \uparrow \downarrow \uparrow \downarrow - + + -  }
\end{equation}
$\Ket{\Phi_{n}^m}$ is therefore a state of size $N= (2n+4)m$, and can be seen as several dipoles oscillating in a sea of pseudo-spins.
Note that $\Ket{\Psi_n^1}$ also belongs to this family when considering periodic boundary conditions.\\

The dimensions and scaling of the Krylov spaces with total system sizes at \emph{fixed} $n$ can also be computed analytically using a transfer matrix approach, for both open and periodic boundary conditions.
The dimension asymptotically scales as $t_+^m$ where $t_+$ is the largest eigenvalue of the transfer matrix $T^{(n)}$ whose entries are given by
\begin{equation}
T_{x, y}^{(n)} = \binom{2n + x-y +1}{n}
\end{equation}
We summarize in Table~\ref{tab:HilbertSpaceScaling-SecondFamily} the asymptotical scaling of the Krylov subspaces dimension for different values of $n$.
Details of the computation of $T^{(n)}$ are kept in App.~\ref{app:secondfamily}.
In particular, for $n=1$, we show that $t_+ = 3 + 2\sqrt{2}$ and that the dimension of the Krylov space scales as
\begin{equation}
\text{dim } \mathcal{K}(\Ket{\Phi_1^m}) \approx \left(t_+ \right)^m \propto 1.342^N \propto 1.158^L.
\end{equation}
In this example, the slow growth of the Krylov space cannot be understood from a simple quasi-particle picture.
Indeed we prove in App.~\ref{app:secondfamily} that there exists no $p \in \mathbb{N}^*$ such that $t_+^p$ is rational.
The subspaces also always scale slower than $2^N$ as can be seen from Table~\ref{tab:HilbertSpaceScaling-SecondFamily} (see also App.~\ref{app:secondfamily}).
Actually, $2^N$ matches the Krylov spaces scaling when $n\rightarrow +\infty$, keeping $m$ fixed.
In the rest of the paper, we focus on the $n=1$ and $n=2$ families, i.e., the families with the two slowest scalings.
We will show in Secs.~\ref{sec:LS-singleK} and \ref{sec:ententropy} that these two families have indeed ergodic statistics at zero and low disorders.

\begin{table}
\begin{tabular}{|c|c|c|c|c|c|}
 \hline n  & 1 & 2 & 3 & 4 & 5\\
 \hline  $\approx$ dim $\mathcal{K}(\Ket{\Phi_n^m}, H)$ & $1.342^N$ & $1.516^N$ & $1.621^N$ & $1.690^N$  & $1.739^N$ \\
\hline
\end{tabular}
\caption{We summarize the scaling of the Krylov subspaces generated by the repeating sequences $\Ket{\left((\uparrow \downarrow)^n - + + -\right)^m}$. The scaling is irrational both in $L$ and $N$ for all values of $n$.
}
\label{tab:HilbertSpaceScaling-SecondFamily}
\end{table}

\section{Level spacing ratio statistics}\label{sec:LS}
Level spacing statistics are a convenient tool to determine whether a system is integrable or ergodic\citep{Wigner1955, Pal2010, Alet2018}.
Random matrices without conservation laws, i.e., describing non-integrable models, have level-repulsion: the probability of having two eigenstates with the same energy is vanishing.
Integrable models, on the other hand, are characterized by the presence of an extensive number of conserved quantities.
Each sector then behaves as an independent random matrix and therefore there is no level repulsion between different sectors.
Additionally, given a symmetry sector, directly studying the level spacing statistics requires unfolding the spectrum.
Indeed, in order to obtain universal signatures, we are required to work with a uniform density of states.
Several unfolding procedures exist, but finite-size effects may lead to different physical interpretations depending on the exact choice of method.\citep{Berry1977, Bruus1997, Gomez2002, Haake2010} 
Instead, an efficient way to characterize quantitatively the level repulsion is to look at the level spacing ratio defined as follow\citep{Oganesyan2007}.
The study of this quantity does not require flattening the density of states.
Let $\{ e_n \}$ be the ordered eigenspectrum of the Hamiltonian. 
We denote by $r_n$ the level spacing ratio
\begin{equation}
r_n = \frac{e_{n+2} - e_{n+1}}{e_{n+1} - e_n}.
\end{equation}
Its probability distribution $P(r)$ distinguishes between ergodic and integrable models.
For an integrable model, $P(r)$ is the Poisson distribution $P_\mathrm{Poi}(r) = \frac{1}{(1+r)^2}$, while for non-integrable systems, it depends on the symmetries of the Hamiltonian and is well-approximated by functionals of the form\citep{Atas2013, Atas2013-2}:
\begin{equation}
P_\beta(r) = \frac{1}{Z_\beta} \frac{(r + r^2)^\beta}{(1 + r + r^2)^{1+\frac{3}{2}\beta}}.
\end{equation}
The real Hermitian Hamiltonians we consider fall into the Gaussian Orthogonal Ensemble (GOE)\citep{Berry1977} with $Z_\beta=\frac{8}{27}$ and $\beta=1$.
%
In practice, it is more convenient to study 
\begin{equation}
\tilde{r}_n = \min ( r_n, \frac{1}{r_n} ) \label{eq:deftilder}
\end{equation} 
which is bounded between $0$ and $1$ and therefore has no heavy tails.
For the classes we are interested in, $P(\tilde{r}) = 2 P(r)\theta(1-r)$.
In the following, references to level ratio are references to $\tilde{r}$.\\

Finally, we remind the reader of the definition of the Kullback-Leibler (KL-)divergence:
\begin{equation}
D_\mathrm{KL} (P, Q) = \int dx p(x) \log (p(x) / q(x) ), \label{eq:defKLDiv}
\end{equation}
where $p$ and $q$ are the probability densities associated to the distributions $P$ and $Q$.
It trivially satisfies $D_\mathrm{KL}(P, P)=0$.
The $KL-$divergence is \emph{asymmetric} in ($P$, $Q$).
It corresponds to the relative entropy from $Q$ to $P$, that is to say the amount of additional information required to model $P$ starting from the prior $Q$.
Hence, when $D_\mathrm{KL}(P_\mathrm{num}(\tilde{r}), P_\mathrm{Poi}(\tilde{r}) ) < D_\mathrm{KL}(P_\mathrm{num}(\tilde{r}), P_\mathrm{GOE}(\tilde{r}) )$, the numerical distribution $P_\mathrm{num}$ is better modelled by the Poisson distribution.

\subsection{Level spacing ratio statistics of the full Hilbert space}
Before we turn to the study of the individual Krylov subspaces themselves in Section \ref{sec:LS-singleK}, we point out that it is crucial to decompose the symmetry resolved Hilbert space into its fractured components, in order to study any thermalization properties and transition.\\

We study a system of length $N=16$  ($L=32$ spins) with OBC in the symmetry sector $p_\mathrm{e} = p_\mathrm{o} = c = 0$ (see Eqs.~\eqref{eq:P-sym} and~\eqref{eq:C-sym}).
The dimension of this symmetry sector is $4.8\times 10^6$, beyond the reach of full diagonalization.
It fractures into approximately $2.5\times 10^5$ Krylov subspaces, whose dimension varies from $1$ to $12870$.
To emphasize the role of the dipole conservation, note that a system of $32$ spins with only the two $U(1)$ sublattice symmetries has already a symmetry sector $(p_\mathrm{e} = 0, p_\mathrm{o} =0)$ that includes $165$M states.\\

\begin{figure}[!]
\subfloat{\includegraphics[width=0.85\linewidth]{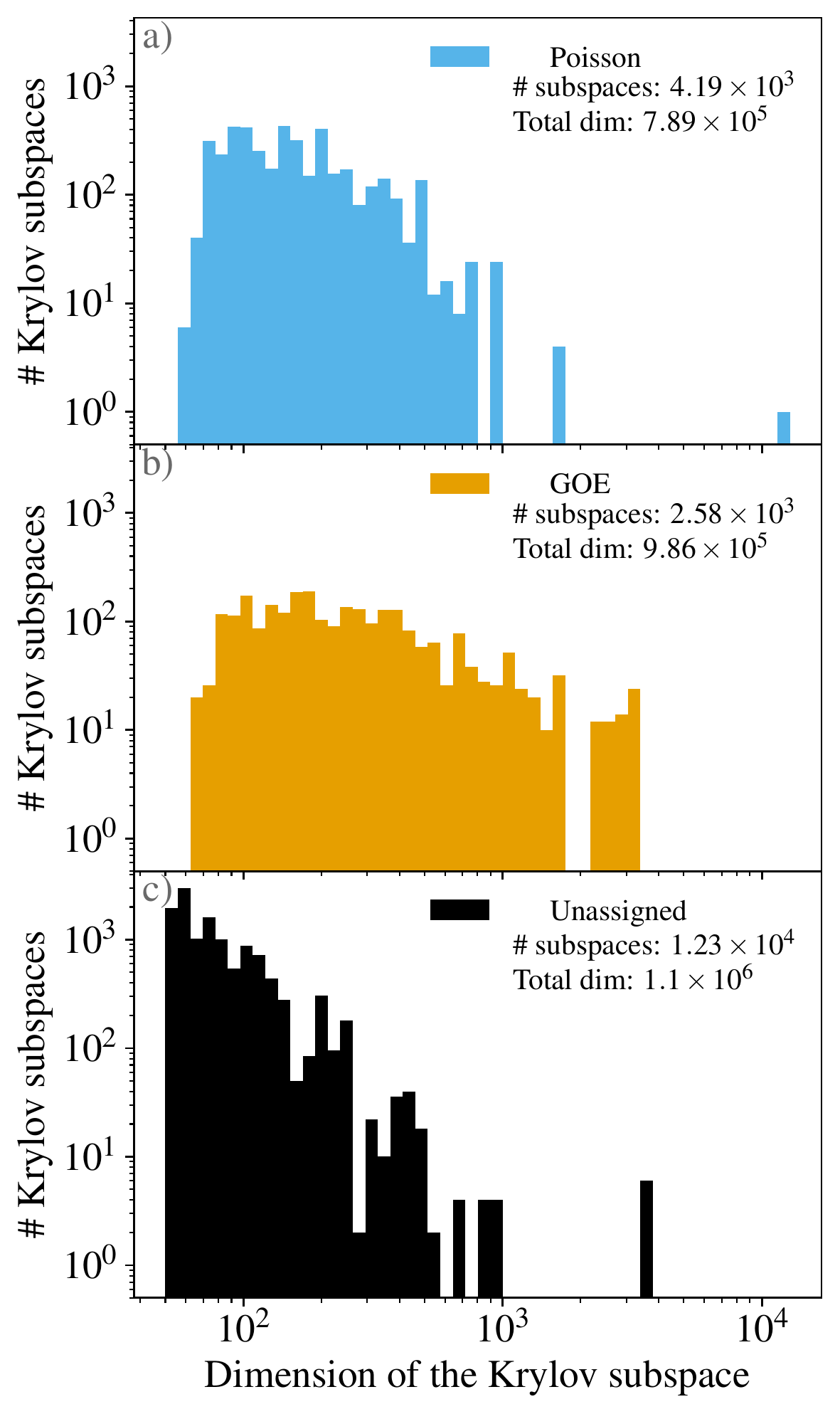}}\\
\subfloat{\includegraphics[width=0.85\linewidth]{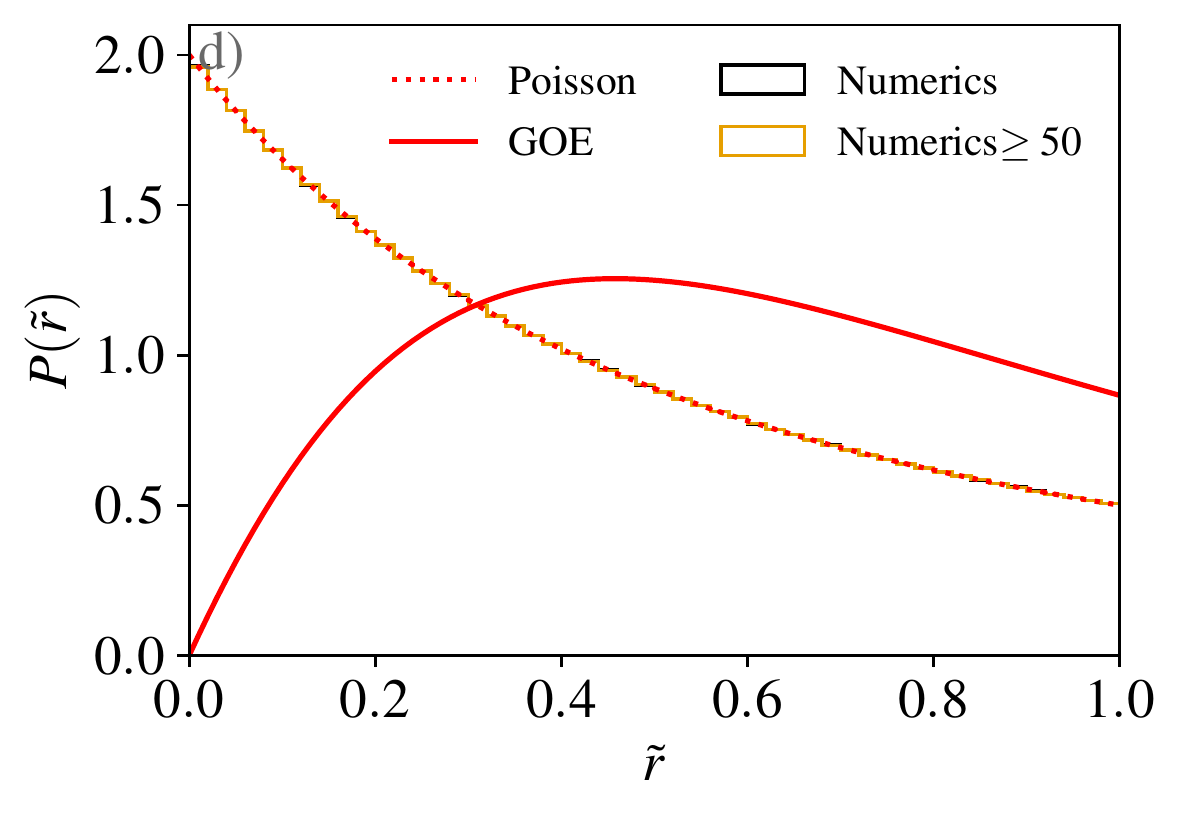}}
\caption{a), b) and c) : histograms of the dimensions of the Krylov subspaces for $N=16$ in the symmetry sector $p_\mathrm{e} = p_\mathrm{o} = c = 0$ depending on their level spacing ratio distribution with $W=0.01$. We only represent subspaces with dimension larger than $50$. In a), we represent the Krylov subspaces that present ergodic level spacing ratio statistics, in b) Poissonian. c) provides the Krylov spaces that cannot be classified by our KL-divergence criteria defined in Eqs.~\eqref{eq:KLcriterium-1} and~\eqref{eq:KLcriterium-2}. In each panel, we also give the number of Krylov spaces in that category, and the total dimension of these Krylov spaces.  d): distribution of the level ratio for the same system if we do not resolve the Krylov spaces. Even if we discard the smallest Krylov subspaces, the level spacing ratio distribution is virtually undistiguishable from the Poisson distribution.}
\label{fig:globalLevelStatistics}
\end{figure}

We compute the exact full spectrum taking advantage of the decomposition into Krylov subspaces and identify whether each Krylov subspace has GOE or Poissonian statistics by computing the KL-divergences defined in Eq.~\eqref{eq:defKLDiv}, using the theoretical distributions as prior.
We fix $W=0.01$ in order to avoid accidental degeneracies and limit finite-size effects, and average over $100$ disorder realizations.
We consider the Krylov subspace to have Poissonian statistics if 
\begin{equation}
D_\mathrm{KL}(P_\mathrm{num}(\tilde{r}), P_\mathrm{Poi}(\tilde{r}) ) < 0.03, \label{eq:KLcriterium-1}
\end{equation} 
and to have GOE statistics if 
\begin{equation}
D_\mathrm{KL}(P_\mathrm{num}(\tilde{r}), P_\mathrm{GOE}(\tilde{r}) ) < 0.03. \label{eq:KLcriterium-2}
\end{equation}
Otherwise, we do not assign a label as the subspace is either afflicted by finite-size effects or presents signs of criticality.
For comparison, the KL-divergences between the Poisson and GOE distributions are given by
\begin{equation}
D_\mathrm{KL}(P_\mathrm{GOE}(\tilde{r}), P_\mathrm{Poi}(\tilde{r}) ) = \frac{1}{3} + \log \frac{\sqrt{3}}{2} \approx 0.189, \label{eq:KL-GOEPoi}
\end{equation}
\begin{equation}
D_\mathrm{KL}(P_\mathrm{Poi}(\tilde{r}), P_\mathrm{GOE}(\tilde{r}) ) = \frac{5 \pi}{2\sqrt{3}} - 3 - \log \frac{27}{8} \approx 0.318. \label{eq:KL-PoiGOE}
\end{equation}
In practice, as we do not compare the distributions directly but histograms with $50$ bins between $0$ and $1$, the effective divergence $ D_\mathrm{KL}(P_\mathrm{Poi}(\tilde{r}), P_\mathrm{GOE}(\tilde{r}) )$ is slightly reduced to $0.305$ [$D_\mathrm{KL}(P_\mathrm{GOE}(\tilde{r}), P_\mathrm{Poi}(\tilde{r}) )$ is almost unaffected].
Our choice of cut-off comes from the following observation: distributions that are maximally confusing with the KL-divergence verify 
\begin{equation}
D_\mathrm{KL}(P_\mathrm{num}(\tilde{r}), P_\mathrm{Poi}(\tilde{r}) ) = D_\mathrm{KL}(P_\mathrm{num}(\tilde{r}), P_\mathrm{GOE}(\tilde{r}) ) \approx 0.05,
\end{equation}
as will be discussed in Sec.~\ref{sec:LS-singleK}.
Choosing a threshold lower than $0.05$ allows us to only select distributions that are convincingly Poissonian or GOE. 
For the sake of simplicity, we also focus only on Krylov spaces of dimension larger than $50$ to minimize the number of samples to average over.
This removes around $2.3\times 10^5$ Krylov spaces associated to $1.8 \times 10^6$ states ($41\%$ of the symmetry sector), including $5 \times 10^4$ dark states, i.e., Krylov spaces consisting of a single state.
The fractions of dark states decreases with system size.
Fig.~\ref{fig:globalLevelStatistics} summarizes our results and the nature of the Krylov spaces. 
Of the approximately $1.9\times 10^4$ remaining spaces, a significant fraction present intermediate statistics (around $1.1\times 10^4$ spaces, comprising $1.1\times 10^6$ states, i.e., $23\%$ of the symmetry sector).
$4.2\times 10^3$ Krylov spaces present clear Poissonian statistics and the remaining $2.6 \times 10^3$ have GOE statistics.
They nonetheless represent a significant proportion of the total symmetry sector, approximately $7.9 \times 10^5$ states ($16 \% $ of the total symmetry sector) and $9.9\times 10^5$ states ($20 \% $) respectively.\\

We now turn towards the study of the level spacing ratio statistics in this symmetry sector, without resolving the Krylov spaces.
As shown in Fig.~\ref{fig:globalLevelStatistics}\href{fig:globalLevelStatistics}{d}, the statistics are essentially undistiguishable from Poisson.
In a given symmetry sector, the occupancies of each Krylov subspace act as an exponential number of additional good quantum numbers.
Therefore there is no apparent level repulsion.
Theoretically, analytical formulas have been recently derived\citep{Sun2020, Giraud2020} to predict the distribution of the level ratios for matrices decomposing in several independent blocks.
These studies computed the level spacing ratio distribution obtained from considering a small number (up to $12$, but easily generalizable) of independent ergodic blocks as a single matrix.
In Ref.~\onlinecite{Giraud2020}, it was numerically shown that the mean level spacing ratio obtained from $M$ ergodic blocks converges toward the Poissonian statistics approximately as $M^{-2}$.
This means that, already for $M=12$, the two average values differ only by $10^{-3}$.
With the exponentially large number of blocks, and the additional scrambling induced by our Poissonian blocks, the precision required to differentiate our numerically obtained distribution from the true Poissonian distribution goes well-beyond any numerically achievable sampling.
Indeed, we numerically obtain that the full numerical distribution $P_\mathrm{full}$, including all the Krylov subspaces, has a KL-divergence with respect to $P_\mathrm{Poi}$ of
\begin{equation}
D_\mathrm{KL} (P_\mathrm{full}(\tilde{r}), P_\mathrm{Poi}(\tilde{r}))\approx 9.7 \times 10^{-8}.
\end{equation}

\subsection{Level spacing ratio statistics in a single Krylov subspace}\label{sec:LS-singleK}
To study the effects of disorder, we focus on the families of Krylov subspaces defined in Sec.~\ref{sec:Krylovsubspaces}.
In particular, we specifically do not consider the largest Krylov subspace.
Indeed, for OBC, the Hamiltonian restricted to this largest Krylov space is equivalent to a random XX model in a transverse field for all system sizes we considered.
It is integrable and localizes at arbitrarily low disorder. 
Let the reduced energy of an eigenstate of energy $E$ be 
\begin{equation}
\varepsilon = \frac{E-E_\mathrm{min}}{E_\mathrm{max}-E_\mathrm{min}},
\end{equation}  with $E_\mathrm{min}$ ($E_\mathrm{max}$) the lowest (highest) energy of the reduced Hamiltonian in the Krylov subspace.
In the rest of this article, we focus on states in the bulk of the spectrum with $\varepsilon \in [0.4,\ 0.6]$.\\

\begin{figure}[ht!]
\subfloat{\includegraphics[width=0.9\linewidth]{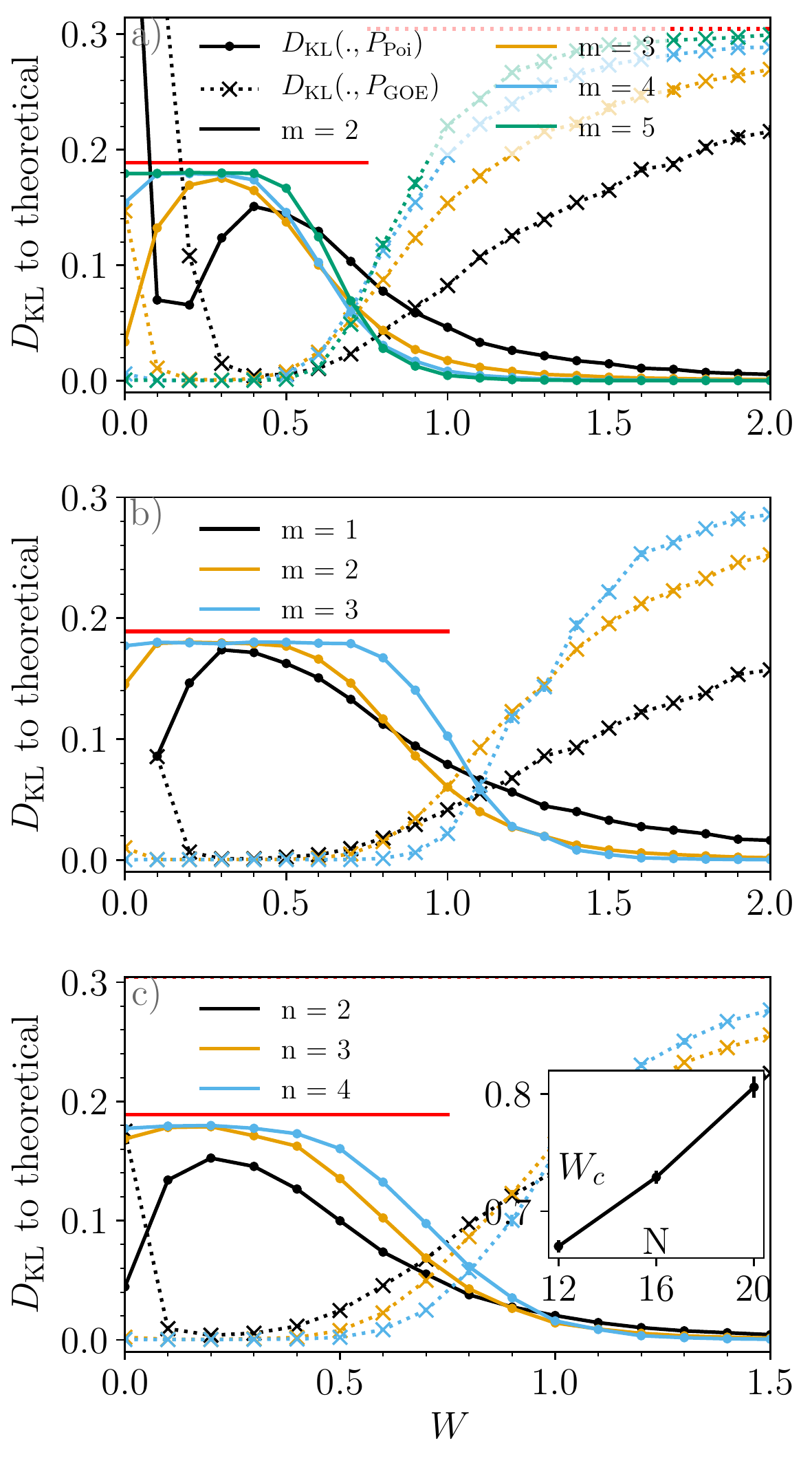}}
\caption{Level spacing ratio statistics for the Krylov subspaces generated by the state $\Ket{\Phi_1^m}$ (a), $\Ket{\Phi_2^m}$ (b) and $\Ket{\Psi_n^1}$ (c) for $\varepsilon \in [0.4\ 0.6]$ and different values of $m$ and $n$.  We compute $D_\mathrm{KL}(P_\mathrm{num}, P_\mathrm{Poi})$ (full line) and $D_\mathrm{KL}(P_\mathrm{num}, P_\mathrm{GOE})$ (dotted line). The red dotted line represent $D_\mathrm{KL}(P_\mathrm{Poi}, P_\mathrm{GOE})$ (see Eq.~\eqref{eq:KL-PoiGOE}) and the full red line $D_\mathrm{KL}(P_\mathrm{GOE}, P_\mathrm{Poi})$ (see Eq.~\eqref{eq:KL-GOEPoi}). For very weak disorder and small Krylov space dimensions, the level spacing ratio distributions present strong finite-size effects due to the sparsity of the model and quasi-degeneracies. Nonetheless, when increasing system sizes, all three families convincingly have ergodic statistics. For all systems, we observe an effective transition from GOE to Poisson statistics. The critical disorder is identified with the crossing points of the two divergences. Note the different effective critical disorder ($W_c\approx 0.75$ for $\Ket{\Phi_1^m}$, $W_c\approx 1.15$ for $\Ket{\Phi_2^m}$) for the first two families with PBC. For the subspaces generated by $\Ket{\Psi_n^1}$ (OBC), on the other hand, we observe an approximately linear shift of the effective critical disorder with increasing system sizes, as is shown in inset. Error bars (generally too small to be seen) are obtained through subsampling of our data.
}
\label{fig:LS}
\end{figure}

We determine the level spacing ratio distribution by averaging over a large number of realizations, ranging from several thousands (for $N=8$) down to $500$ for the larger systems (for $N=30$).
We represent in Fig.~\ref{fig:LS}\href{fig:Ls}{a-c} the KL divergence of the distribution of the level spacing ratio in the three families of Krylov subspaces defined in Secs~\ref{sec:Krylov-basic} and \ref{sec:Krylov-slow}, using GOE and Poisson distributions as prior.
For all families, at low disorder, we observe a quick convergence towards the GOE distribution of the level spacing ratio distribution, when increasing the system size $N$.
The three families of Krylov subspaces appear indeed ergodic in the thermodynamic limit.
Crossing of the KL-divergence universally occurs for $D_\mathrm{KL}(P_\mathrm{num}(\tilde{r}), P_\mathrm{Poi}(\tilde{r}) ) \approx 0.05$.
This implies that we are maximally confused about which theoretical distribution better approximates the numerical one at this value of the KL-divergence.
Thus, we take this crossing as a marker of the phase transition.\\

We first turn to the Krylov spaces generated by $\Ket{\Phi^m_1}$ defined in Eq.~\eqref{eq:Secondfamily}, working with periodic boundary conditions due to the favorable scaling.
As shown in Fig.~\ref{fig:LS}\href{fig:Ls}{a}, for $m>2$, the crossing point of the KL divergences shows very small finite-size effects at $W_c^1\approx 0.75$ despite the small Hilbert spaces.
We have performed the same analysis by studying the evolution of the mean level spacing ratio (see Appendix \ref{app:AdditionalData_meanlevel}). 
For the second family generated by $\Ket{\Phi_2^m}$, also with PBC, we observe similar results in Fig.~\ref{fig:LS}\href{fig:Ls}{b}.
Due to the faster growth of the Krylov subspaces, we are effectively limited to smaller systems plagued by stronger finite-size effects.
For each $m$, we observe a transition from an ergodic phase to a localized phase, albeit at a significantly larger disorder strength, despite the similar Hilbert space dimensions and structures.
The effective critical disorder strengths do not display any simple convergence behavior when increasing $m$, at least within the accessible system sizes.
Finally, the family $\Ket{\Psi_n^1}$ --- studied with OBC due to the slower scaling --- also exhibits signs of an MBL phase transition, as shown in Fig.~\ref{fig:LS}\href{fig:LS}{c}.
Interestingly enough, the crossing point admits an approximately linear drift with increasing system sizes (see inset in Fig.~\ref{fig:LS}\href{fig:LS}{c}).
Additionally, we observe in all Krylov subspaces that the critical disorder strength strongly depends on the relative energies of the eigenstates.
Mobility edges\citep{Gornyi2005, Basko2006, Luitz2015} are therefore also present in these constrained systems.
More details can be found in App.~\ref{app:AdditionalData_mobedge}.\\

Note that the three Krylov spaces considered here originate from the same initial Hamiltonian (up to boundary conditions) and therefore for a given $W$ and $N$ have the same disorder and hopping amplitudes in configuration space.  
Still, the corresponding critical values, as predicted by the level spacing ratio distributions, and scaling behavior are \emph{radically} different.
$\mathcal{K}(\Ket{\Phi_2^3})$ and $\mathcal{K}(\Ket{\Phi_1^4})$ both correspond to $N=24$, $\mathcal{K}(\Ket{\Psi_{3}^1})$ and $\mathcal{K}(\Ket{\Phi_2^2})$ to $N=16$, and $\mathcal{K}(\Ket{\Psi_{4}^1})$ and $\mathcal{K}(\Ket{\Phi_1^2})$ to $N=12$ and yet admit different transition points.
Conversely, the Krylov space dimension alone is also not a good indicator of the critical disorder: both $\mathcal{K}(\Ket{\Phi_2^2})$ ($N=16$) and $\mathcal{K}(\Ket{\Psi_{4}^1})$ ($N=20$) have a dimension close to $2.2\times 10^{4}$.
$\mathcal{K}(\Ket{\Phi_2^2})$ appears to localize at a larger disorder than $\mathcal{K}(\Ket{\Psi_{4}^1})$ even though the disorder in the Fock basis is averaged over less sites.

The family $\mathcal{K}( \Ket{\Psi_n^1})$ shows a strong drift of the critical disorder towards larger values with increasing system sizes.
This could be a sign of an absence of a transition for these subspaces in the thermodynamic limit.
Paradoxically, this family also has a structure very close to an integrable one.
Indeed, the Hamiltonian acting on the sea of pseudo-spins $\Ket{\uparrow}$ and $\Ket{\downarrow}$ reduces to a non-interacting XX Hamiltonian.
The pair of dipoles breaks integrability by stitching together a set of triplets---the sea of pseudo-spins to the left, in-between, and to the right of the pair of dipoles---of integrable spaces.
Yet, while the XX Hamiltonian is localized at arbitrarily low-disorder, with an effective critical disorder strength decreasing with system size, we observe the exact opposite for $\Ket{\Psi_n^1}$.

The different behavior observed in our Krylov subspaces reinforces the need to distinguish between the Krylov subspaces if we want to study the MBL phase transition and the effect of disorder.
In App.~\ref{app:AdditionalData_bothStats}, we show some additional numerical results showing the level spacing ratio statistics obtained when mixing the subspaces generated by $\Ket{\Phi^3_2}$ and $\Ket{\Phi^4_1}$.
We observe a significant difference between the distribution at low-disorder and $P_\mathrm{GOE}$, and a smoother crossover when studying the KL-divergences.

\section{Entanglement entropy in a constrained model}\label{sec:ententropy}
In the previous Section, we have seen that the different Krylov subspaces appear to undergo an MBL phase transition at different critical disorder strengths, according to their level spacing ratio distributions.
We now turn to the study of the von-Neumann entanglement entropy (vNEE) of the many-body eigenstates as another complementary probe of this transition.
For a subsystem $\mathcal{A}$, the vNEE of the pure state $\Ket{\Psi}$ is given by:
\begin{equation}
S(\mathcal{A}) = - \text{Tr}(\rho_{\mathcal{A}} \log \rho_{\mathcal{A}} ) \text{ with } \rho_{\mathcal{A}} = \text{Tr}_{\nA} \Ket{\Psi}\Bra{\Psi},
\end{equation}
where $\text{Tr}_{\nA}$ marks the trace on the degrees of freedom not in $\mathcal{A}$.
We denote by $\mathcal{H}_\Aa$ (resp. $\mathcal{H}_{\nA}$) the Hilbert space of $\rho_\Aa$ (resp. $\rho_{\nA}$).
In terms of the entanglement entropy, the MBL phase transition can be seen as a transition from thermal volume-law to an area-law\citep{Bauer2013, Kjall2014, Luitz:2016hr,Yu:2016gb}.
In one dimension, the volume law is to be understood as
\begin{equation}
S(\mathcal{\Aa}) \propto s_\mathrm{th}\, l_\Aa \text{ for }l_\Aa \ll L
\end{equation}
with $l_\Aa$ the number of sites (degrees of freedom) in $\Aa$. 
$s_\mathrm{th}$ takes the value $\log 2$ in the thermal phase for conventional spin-$\frac{1}{2}$ systems.
The strongly constrained model we study sees very irregular growth of the Hilbert space $\mathcal{H}_\Aa$ with subsystem size.
Instead, we consider the entanglement entropy to be ergodic if it verifies:
\begin{equation}
S(\mathcal{\Aa}) \approx S_\mathrm{Page}(\mathcal{\Aa}),
\end{equation}
where the Page entropy\citep{Page1993} $S_\mathrm{Page}$ is the average entanglement entropy of uniformly distributed random states.
In the absence of symmetries or of Hilbert space fragmentation, the Page entropy $S_\mathrm{Page}(\mathcal{\Aa})$ is given by
\begin{equation}
 S_\mathrm{Page}(\mathcal{\Aa}) \approx \log m - \frac{m}{2 M} \text{ for } 1 \leq m < M,
\end{equation}
\begin{align}
\text{ with }  m &= \min(\dim \mathcal{H}_\Aa,\dim  \mathcal{H}_{\nA}),\label{eq:defm} \\
 M &= \max(\dim  \mathcal{H}_\Aa,\dim  \mathcal{H}_{\nA}).
\end{align} 
The Page entropy trivially satisfies the volume law as $\log m$ is roughly proportional to the number of degrees of freedom in $\Aa$.
Due to the presence of the multiple $U(1)$ symmetries, we have to take into account the splitting of the wave functions down to submatrices in different symmetry subsectors.
Correspondingly, the Hilbert space (or Krylov subspace) can be split into:
\begin{align}
\mathcal{H} = \bigoplus_j \mathcal{H}_{\Aa, j} \otimes \mathcal{H}_{\nA, j}
\end{align}
where the subspaces $\mathcal{H}_{\Aa, j}$ and $\mathcal{H}_{\nA, j}$ have dimension
\begin{align}
m_j &= \min (\dim \mathcal{H}_{\Aa, j}, \dim \mathcal{H}_{\nA, j} ), \label{eq:defmj}\\
M_j &= \max (\dim \mathcal{H}_{\Aa, j}, \dim \mathcal{H}_{\nA, j} ), \label{eq:defMj}
\end{align}
such that\footnote{In principle, due to constraints not directly taken into account by the symmetries, it is possible that $\mathrm{dim}\mathcal{H} < \sum\limits_j m_j M_j$. Then, our generalized formula in Eq.~\eqref{eq:GeneralizedPage} is no longer valid. It is not the case in our model.}
$\mathrm{dim}\mathcal{H} = \sum\limits_j m_j M_j$.
The Page entropy is then given by
\begin{multline}
S_\mathrm{Page}(\mathcal{\Aa}) = \sum\limits_j \left[ \frac{m_j M_j}{\mathrm{dim} \mathcal{H}} \left(\log m_j - \frac{m_j}{2 M_j} \right) \right.\\
 \left. - \frac{m_j M_j}{\mathrm{dim} \mathcal{H}} \log \frac{m_j M_j}{\mathrm{dim} \mathcal{H}} \right]\label{eq:GeneralizedPage}
\end{multline}
The area-law remains here defined as
\begin{equation}
S(\mathcal{\Aa}) = O(1) \text{ when } l_\Aa,\, L \rightarrow + \infty
\end{equation}

\begin{figure*}[ht!]
\subfloat{\includegraphics[width=\linewidth]{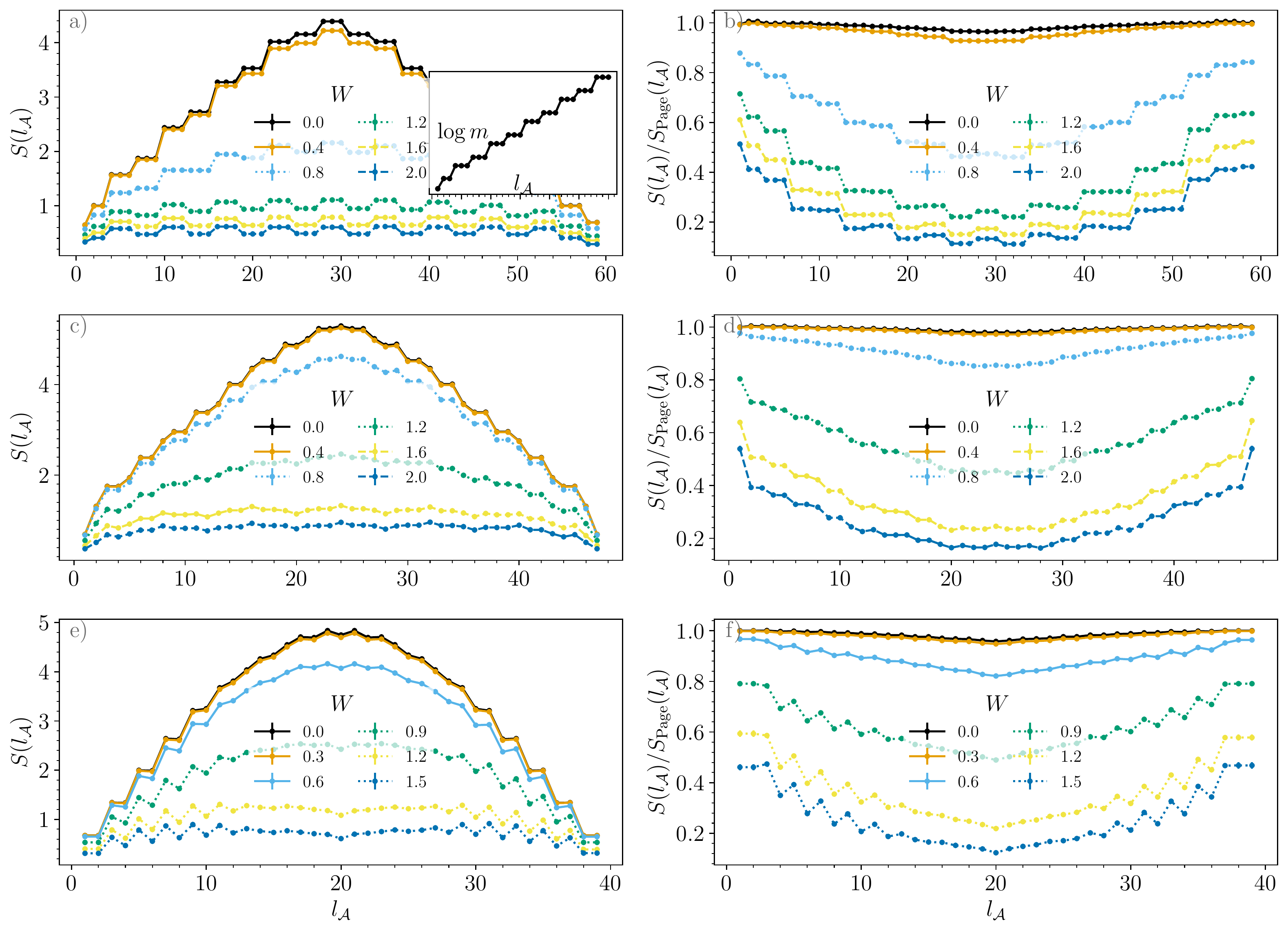}}
\caption{Entanglement entropy as function of the subsystem size for several disorder strengths $W$ for $\Ket{\Phi_1^5}$ (a-b), $\Ket{\Phi_2^3}$ (c-d) and $\Ket{\Psi_4^1}$ (e-f). a), c) and e): vNEE. b), d) and f): vNEE normalized by the Page entropy given in Eq.~\eqref{eq:GeneralizedPage}. For all Krylov subspaces, we observe a clear transition between a volume-law at low-disorder, where the entropy varies little with disorder strength. The slight dip in the middle of the chain for the normalized vNEE is characteristic of finite-size systems\citep{Page1993} as $m_j$ and $M_j$ in Eqs.~\eqref{eq:defmj} and \eqref{eq:defMj} become of the same order. It is unrelated to any breakdown of the volume law. At stronger disorder, the vNEE transitions towards an area law. In the inset in a), we display $\log m$ as defined in Eq.~\eqref{eq:defm} for $l_\mathcal{A}$ in $[\![1, 30]\!]$. The irregular growth explains the pattern seen in the entropy in the ergodic and MBL phases. Error bars are too small to be seen.
}
\label{fig:entanglementGrowth}
\end{figure*}

We compute the entanglement entropy in the different Krylov spaces introduced in Secs.~\ref{sec:Krylov-basic} and \ref{sec:Krylov-slow}, and average over all states with $\varepsilon \in [0.4, 0.6]$ and over a large number of disorder realizations (see Sec.~\ref{sec:LS-singleK}).
We work in the original spin basis ($\Ket{0}$, $\Ket{1}$).
We assume PBC for $\Ket{\Phi_m^1}$ and $\Ket{\Phi_m^2}$, and OBC for $\Ket{\Psi_n^1}$ due to the favorable scalings.
In Fig.~\ref{fig:entanglementGrowth}, we show the scaling of the vNEE $S(l_\mathcal{A})$ as a function of the subsystem size, where $\Aa$ is the segment made of  the $l_\Aa$ consecutive spins $[\![1, l_\Aa]\!]$ for different disorder values for $\Ket{\Phi^5_1}$, $\Ket{\Phi^3_2}$ and $\Ket{\Psi^1_4}$.
The entropy we obtain therefore matches the one we would obtain in the pseudo-spins basis when $l_\Aa$ is even.
At low disorder values, the vNEE remains roughly proportional to the Page entropy, following the aforementioned volume law.
The entropy varies only weakly with the disorder strength.
At stronger disorder, we observe a crossover towards an area law where the entanglement remains (nearly) constant over several decades. 
This area law is typical of the predicted MBL phase and shows no signs of increasing again at larger scales.\\

The exact pattern followed by the vNEE depends on the Krylov subspaces, and can be very irregular.
In particular, the family $\Ket{\Phi^m_1}$, for the cut we chose, has $S(3l+1)=S(3l+2)=S(3l+3)$ for $l \geq 1$.
It is not a consequence of any effective three-spin quasi-particles but a non-trivial interplay between the pair-hopping terms and the chosen starting state (see App.~\ref{app:redmat}).
Additionally, as can be seen in Fig.~\ref{fig:entanglementGrowth}\href{fig:entanglementGrowth}{a}, in the ergodic phase, the growth of the entanglement entropy from one plateau to the next alternates between large and small jumps.
This irregular growth pattern comes from the lack of translation invariance at the single spin level in the starting generating state (while it remains invariant by translation of $12$ spins).
The dimension of the reduced density matrix grows faster when $l_\Aa$ goes through a higher entropy jump.
This irregular growth also affects the MBL phase.
A larger growth of the reduced Hilbert space translates into more states connected by a pair-hopping term going through the entanglement cut.
As the entanglement entropy at large disorder mainly arises from local resonant pairs, this structure leads to the observed alternating high and low plateaus.
More details on the growth of the dimension of the reduced density matrix and the pairing structure can be found in App.~\ref{app:redmat}).\\

\begin{figure}[ht!]
\includegraphics[width=0.9\linewidth]{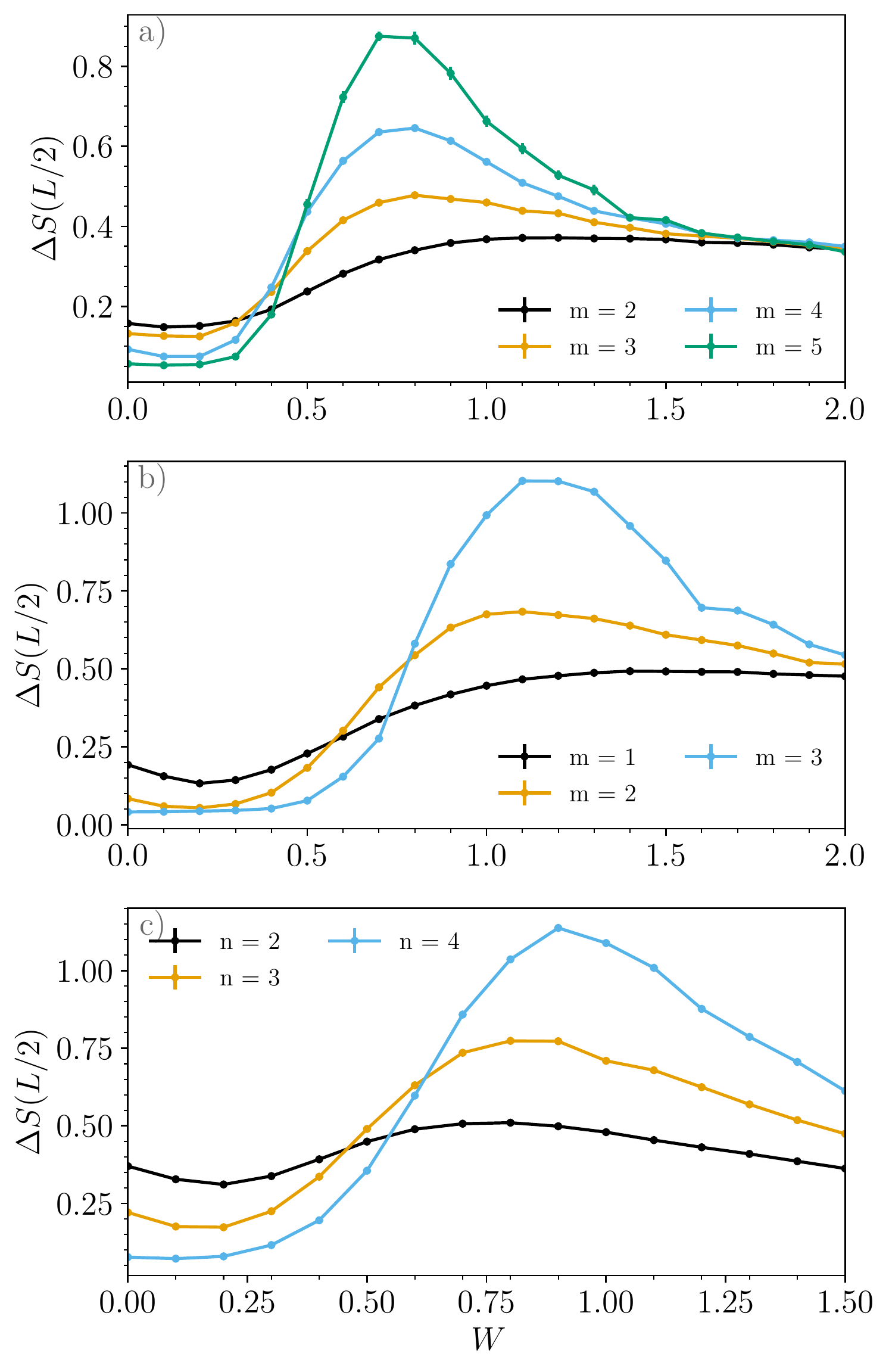}
\caption{Standard deviation of the mid-chain entanglement entropy for $\Ket{\Phi^m_1}$ (a), $\Ket{\Phi_2^m}$ (b) and $\Ket{\Psi_n^1}$ for different $m$ and $n$ as a function of the disorder strength $W$. 
The maximum of the standard deviation is supposed to capture the phase transition. For $\Ket{\Phi_1^m}$ (a), we observe a clear and more and more marked peak around $W=0.75$. For $\Ket{\Phi^m_2}$ (b), it is harder to extract a tendency within the system size available, due to a strong finite size effect. Finally, for $\Ket{\Psi_n^1}$ (c), we observe an approximately linear increase of the effective critical disorder with system sizes. 
In all cases, the predictions qualitatively agree with the ones obtained from the level spacing ratio statistics. Error bars are too small to be seen.}
\label{fig:entanglementSTD}
\end{figure}

To pinpoint the transition, it is convenient to study the standard deviation of the entanglement entropy (typically at the midchain point)\citep{Kjall2014}.
The transition point is taken to be at its maxima: the system can there be either in a thermal state with high volume-law entanglement or in a localized states with low entanglement.
In Fig.~\ref{fig:entanglementSTD}, we show the standard deviation of the entanglement entropies obtained for all states with $\varepsilon \in [0.4, 0.6]$ and for different disorder realisations for the Krylov subspace we considered.
For all families, the larger the system, the more peaked the standard deviation is.
For the family generated by $\Ket{\Phi_1^m}$, the peaks clearly concentrate around the critical disorder value $W_c \approx 0.75$. 
For $\Ket{\Phi^m_2}$ there is no clear tendency emerging. 
Finally, for $\Ket{\Psi_n^1}$, the effective critical disorder values increase quasi-linearly with system size, preventing pinpointing any phase transition.
The obtained values are in qualitative agreement with those obtained considering the level spacing ratio. 
Due to the limited number of sizes available in each family, we cannot perform a reliable scaling analysis.

\section{Discussions and conclusions}
\begin{figure}[ht!]
\begin{center}
\subfloat{\includegraphics[width=\linewidth]{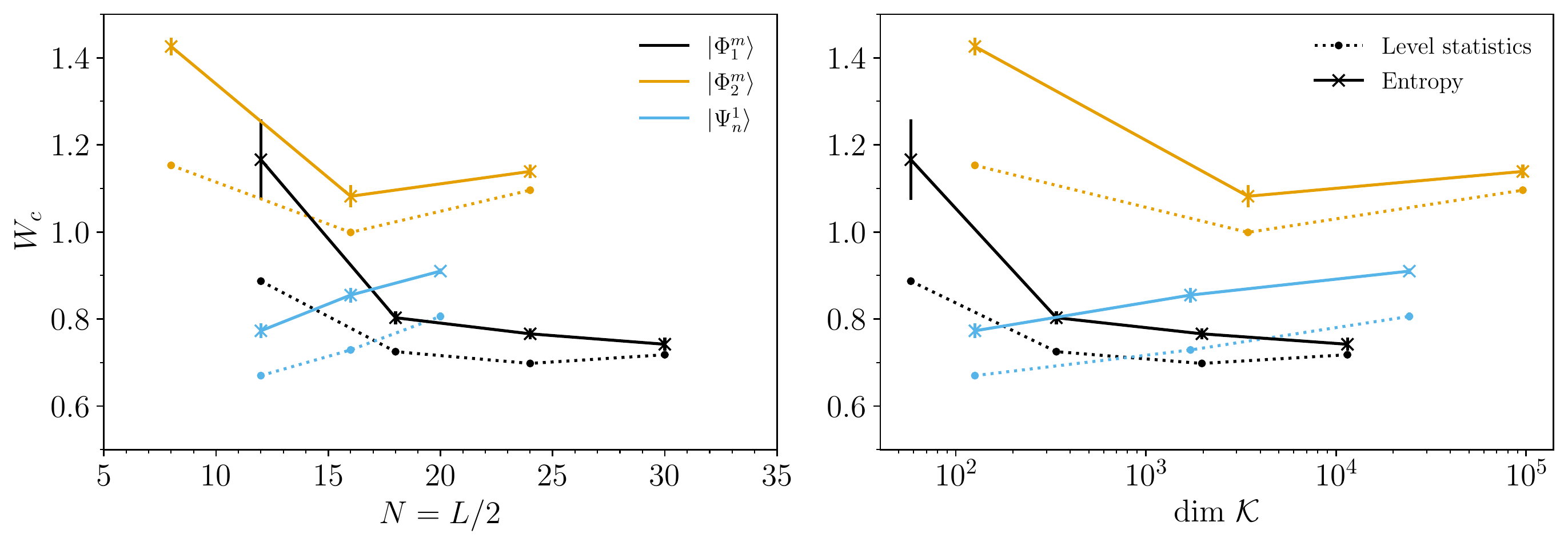}}
\end{center}
\caption{Critical disorder strengths obtained from the study of the level spacing ratio distributions and of the entanglement entropy for the different Krylov spaces we investigated in this paper. No distinct pattern emerges relating system sizes (a) or dimensions of the Krylov subspaces (b) to the critical disorder strengths. Error bars for entanglement are obtained by fitting randomly generated sequences with similar properties of our data and measuring the variation of the interpolated maximum.
}
\label{fig:PD}
\end{figure}

In this paper, we have provided numerical evidence of a many-body localisation type transition within the ergodic Krylov subspaces of constrained models presenting a strong fragmentation of the Hilbert space. 
Due to the slow scaling of the Hilbert space dimensions, we have been able to study systems comprised of up to $60$ spins using exact diagonalization.
We observe the transition from an approximate linear scaling of the entanglement entropy at low-disorder to a clear area law over significantly larger scales than conventionally studied.
We see no signs of a general breakdown of the many-body localization phenomenon in these large systems, though the Krylov spaces' dimensions remain comparable to other models which have been studied.
Within the same constrained model, the different Krylov subspaces see a transition occuring at wildly different disorder strengths.
This reinforces the importance of considering separately each Krylov space to study the localization properties of systems that see such a fragmentation of the full Hilbert space: without doing so, any sign of the transition will be blurred towards Poissonian statistics.
More importantly, we see no significant correlations between effective critical disorder strengths and Krylov space dimensions or system sizes, as illustrated in Fig.~\ref{fig:PD}.

The role of the structure of the Krylov space is therefore key to explaining the MBL phase transition in these models, and a detailed study is left for future works.
The subspaces generated by the families $\Ket{\Phi_1^m}$ appear to present a stable MBL phase transition in the thermodynamic limit, whether we consider the level spacing ratio distributions or the entropy properties. 
On the other hand, the subspaces generated by $\Ket{\Psi^1_n}$ show an approximately linear scaling of the critical disorder strengths with system size, within the sizes and the numerical precision we have access to.
It raises the question whether this subspace actually always thermalizes in the thermodynamic limit.
This is especially remarkable given that this subspace and the action of the constrained Hamiltonian on it appear the closest to an effective integrable XX model given the presence of a single pair of dipoles in a sea of integrable spins.

\acknowledgements
We thank David Aceituno, Fabien Alet, Jeremy Bensadon, Marta Brzezi\'{n}ska, Vardan Kaladzhyan, Nicolas Laflorencie and Nicolas Macé for useful discussions.
N.R. is also grateful to B.A. Bernevig and S. Moudgalya for collaboration on previous related works.
L.H and J.B. were supported by the ERC Starting Grant No.~679722, the Roland Gustafsson's Foundation for Theoretical Physics and the Karl Engvers foundation.
N.R. was supported by NSF through the Princeton University’s Materials Research Science and Engineering Center DMR-2011750B, the DOE Grant No. DE-SC0016239, the Schmidt Fund for Innovative Research, Simons Investigator Grant No. 404513, the Packard Foundation, the NSF-EAGER No. DMR 1643312, NSF-MRSEC No. DMR-1420541 and DMR-2011750, ONR No. N00014-20-1-2303, Gordon and Betty Moore Foundation through Grant GBMF8685 towards the Princeton theory program, BSF Israel US foundation No. 2018226, and the Princeton Global Network Funds.

\appendix

\section{Computation of the dimension of the Krylov subspaces}

\subsection{Krylov subspaces generated by $\Ket{\Psi_n^1}$}\label{app:Krylovinfty}
\begin{table*}
\begin{tabular}{|c|c|c|c|c|c|c|c|}
 \hline n & N  & $\dim \mathcal{K}(\Ket{\Psi_n^1}, H_\mathrm{OBC})$ & $\mathcal{C}(\mathcal{H}^\mathrm{OBC})$ & $\dim \rho_{\frac{1}{2}}^\mathrm{OBC}$ & $\dim \mathcal{K}(\Ket{\Psi_n^1}, H_\mathrm{PBC})$ & $\mathcal{C}(\mathcal{H}^\mathrm{PBC})$ & $\dim \rho_{\frac{1}{2}}^\mathrm{PBC}$  \\
\hline 2 & 12 & 126 & 4.75 & 16 & 630 & 5.33 & 80\\
\hline 3 & 16 & 1716 & 6.77 & 64 & 12012 & 7.38 & 448\\
\hline 4 & 20 & 24310 & 8.8 & 256 & 218790 & 9.41 & 2304\\
\hline 5 & 24 & 352716 & 10.82 & 1024 & - & - & - \\
\hline
\end{tabular}
\caption{We summarize the properties of the Krylov space for the family generated by $\Ket{\Psi^1_n}$. The columns $3$ to $5$ ($6$ to $8$) are for the OBC (PBC) case. The third and sixth columns list the dimensions of the Krylov subspaces according to Eqs.~\eqref{eq:dimPsiPBC} and \eqref{eq:dimPsiOBC}. The dimension of the Krylov spaces approximately grows as $2^N$. The fourth and seventh columns show the connectivity of the Hamiltonian $\mathcal{C}$ (here defined as the ratio of the number of non-zero non-diagonal terms in the Hamiltonian over the Hilbert space dimension). Finally, the fifth and eighth columns list the dimensions  $\dim \rho_{\frac{1}{2}}^\mathrm{OBC}$ of the reduced density matrix for a cut exactly in the middle of the chain, i.e., separating the two $+$s in the generating state for different system sizes. }
\label{tab:HilbertSpaceDim-Firstvariant}
\end{table*}

We discuss in this Appendix the simplification of the formulas given in Eqs.~\eqref{eq:dimPsiPBC} and \eqref{eq:dimPsiOBC}.
We start with periodic boundary conditions, where the dimension of the Krylov subspace is given by:
\begin{equation}
\text{dim }\mathcal{K}(\Ket{\Psi_{n}^1})_{\mathrm{PBC}} = \sum\limits_{n_\uparrow, n_\downarrow=0}^{N_\uparrow}  \binom{2N_\uparrow - n_\uparrow - n_\downarrow}{N_\uparrow-n_\uparrow} \binom{n_\uparrow + n_\downarrow}{n_\downarrow}.
\end{equation}
We reorganize the double summation introducing $s=n_\uparrow + n_\downarrow$,
\begin{equation}
\text{dim }\mathcal{K}(\Ket{\Psi_{n}^1})_{\mathrm{PBC}} = \sum\limits_{s=0}^{2 N_\uparrow} \sum\limits_{n_\uparrow=\max (0, s-N_\uparrow)}^{\min(N_\uparrow, s)}  \binom{2N_\uparrow - s}{N_\uparrow-n_\uparrow} \binom{s}{n_\uparrow}.
\end{equation}
The bounds of the sum on $n_\uparrow$ can be simplified as either one of the binomial coefficient is $0$ for $n_\uparrow < \max (0, s-N_\uparrow)$ or $n_\uparrow > \min(N_\uparrow, s)$.
Namely, we get
\begin{equation}
\text{dim }\mathcal{K}(\Ket{\Psi_{n}^1})_{\mathrm{PBC}} = \sum\limits_{s=0}^{2 N_\uparrow} \sum\limits_{n_\uparrow=0}^{N_\uparrow}  \binom{2N_\uparrow - s}{N_\uparrow-n_\uparrow} \binom{s}{n_\uparrow}.
\end{equation}
From here, application of the Chu-Vandermonde identity 
\begin{equation}
\sum\limits_{j=0}^k \binom{m}{j}\binom{n}{k-j} = \binom{n+m}{k}
\end{equation}
leads to
\begin{equation}
\text{dim }\mathcal{K}(\Ket{\Psi_{n}^1})_{\mathrm{PBC}} = \sum\limits_{s=0}^{2 N_\uparrow} \binom{2N_\uparrow}{N_\uparrow}  = \frac{(2N_\uparrow+1)!}{N_\uparrow!^2}.
\end{equation}

We now turn to open boundary conditions. The dimension of the Krylov space (denoted here $d_{\mathrm{OBC}}$ for convenience) is given by
\begin{widetext}
\begin{align}
d_{\mathrm{OBC}} &= \sum\limits_{n_\uparrow, n_\downarrow=0}^{N^L_\uparrow} \binom{2N^L_\uparrow-n_\uparrow}{N^L_\uparrow} \binom{2N^L_\uparrow-n_\downarrow}{N_\uparrow^L} \binom{n_\uparrow + n_\downarrow}{n_\downarrow}\\
&= \sum\limits_{s=0}^{2 N^L_\uparrow} \sum\limits_{n_\uparrow=\max (0, s-N^L_\uparrow)}^{\min(N^L_\uparrow, s)}  \binom{2N^L_\uparrow-n_\uparrow}{N^L_\uparrow} \binom{2N^L_\uparrow+n_\uparrow - s}{N^L_\uparrow} \binom{s}{n_\uparrow}\\
&= \sum\limits_{s=0}^{2 N^L_\uparrow} \sum\limits_{n_\uparrow=0}^{N^L_\uparrow}  \binom{2N^L_\uparrow-n_\uparrow}{N^L_\uparrow} \binom{2N^L_\uparrow+n_\uparrow - s}{N_\uparrow^L} \binom{s}{n_\uparrow}\\
&=\sum\limits_{n_\uparrow=0}^{N^L_\uparrow}  \binom{2N^L_\uparrow-n_\uparrow}{N^L_\uparrow}\left(\sum\limits_{s=0}^{2N_\uparrow} \binom{2N^L_\uparrow+n_\uparrow - s}{N_\uparrow^L} \binom{s}{n_\uparrow} \right)
\end{align}
\end{widetext}
Using the Chu-Vandermonde identity
\begin{equation}
\sum\limits_{m=0}^n \binom{m}{j}\binom{n-m}{k-j} = \binom{n+1}{k+1},\label{eq:chuVand2}
\end{equation}
and the fact that the term in the second sum is zero for $s\geq 2 N^L_\uparrow + 1$, we obtain
\begin{equation}
d_{\mathrm{OBC}} = \sum\limits_{n_\uparrow=0}^{N^L_\uparrow}  \binom{2N^L_\uparrow-n_\uparrow}{N^L_\uparrow}\binom{2N^L_\uparrow+n_\uparrow + 1}{N_\uparrow^L}.\label{eq:appTemp}
\end{equation}
Now we introduce $t=2N^L_\uparrow - n_\uparrow$ such that
\begin{align}
d_{\mathrm{OBC}} &= \sum\limits_{t=N_\uparrow}^{2N_\uparrow^L}  \binom{t}{N^L_\uparrow}\binom{4N^L_\uparrow + 1 - t}{N_\uparrow^L}\\
&= \sum\limits_{t=0}^{2N^L_\uparrow}  \binom{t}{N^L_\uparrow}\binom{4N^L_\uparrow + 1 - t}{N_\uparrow^L}.
\end{align}
Similarly, we can instead introduce $\tilde{t} = 2N^L_\uparrow + n_\uparrow +1$ to obtain
\begin{align}
d_{\mathrm{OBC}} &= \sum\limits_{\tilde{t}=2N_\uparrow + 1}^{3N_\uparrow^L +1} \binom{4N^L_\uparrow + 1 - \tilde{t}}{N_\uparrow^L} \binom{\tilde{t}}{N^L_\uparrow}\\
&= \sum\limits_{\tilde{t}=2N_\uparrow + 1}^{4N_\uparrow^L +1}   \binom{\tilde{t}}{N^L_\uparrow}\binom{4N^L_\uparrow + 1 - \tilde{t}}{N_\uparrow^L}.
\end{align}
This leaves us with
\begin{align}
d_{\mathrm{OBC}} &=\frac{1}{2} \sum\limits_{t=0}^{4N_\uparrow^L +1} \binom{4N^L_\uparrow + 1 - t}{N_\uparrow^L} \binom{t}{N^L_\uparrow}\\
 &=\frac{1}{2} \binom{4N_\uparrow^L +2}{2N_\uparrow^L +1} = \binom{4N_\uparrow^L +1}{2N_\uparrow^L},
\end{align}
where we used Eq.~\eqref{eq:chuVand2} a second time and obtain  Eq.~\eqref{eq:dimPsiOBC2} in the main text. 
Note that as far as we know, this special identity for the triple sum of binomials is not registered in conventional tables.
It can be generalized to
\begin{multline}
d(X, Y) = \sum\limits_{x, y =0}^{X, Y} \binom{2X-x}{X} \binom{2Y-y}{Y} \binom{x+y}{x} \\
= \binom{2X +2Y+1}{X+Y},
\end{multline}
where we used, following Eq.~\eqref{eq:appTemp},
\begin{align}
d(X, Y) &= \sum\limits_{x=0}^{2X}  \binom{2X-x}{X}\binom{2Y +x + 1}{Y},\\
d(X, Y) &= \sum\limits_{y=0}^{2Y}  \binom{2Y-y}{Y}\binom{2X +y + 1}{X},
\end{align}
the two changes of variables $t_x = 2X-x$ and $t_y = 2Y + y + 1$, and applied Eq.~\eqref{eq:chuVand2}.

\subsection{Krylov subspaces generated by $\Ket{\Phi_n^m}$}\label{app:secondfamily}
We now turn to the asymptotic dimension scaling of the Krylov spaces generated by $\Ket{\Phi_n^m}$.
Let us first consider periodic boundary conditions for simplicity.
$\uparrow$'s can move in between the two dipoles to their left (but not beyond), and similarly for $\downarrow$'s to their right.
We denote $x_j=0, ..., n$ and $y_j=0, ..., n$ the number of pseudo-spins of the $j^\text{th}$ sequence of pseudo spins that have moved to their left and to their right.
For example, we consider the Krylov subspace generated by $\Ket{\Phi_2^3}$, i.e.,
\begin{equation}
\vert \uparrow \downarrow \uparrow \downarrow - ++- \uparrow \downarrow \uparrow \downarrow - ++- \uparrow \downarrow \uparrow \downarrow - ++- \rangle
\end{equation}
A typical configuration connected to this initial state looks like
\begin{equation}
\lvert \uparrow\uparrow \downarrow - + \underbrace{\uparrow \downarrow \uparrow}_{
\scriptsize{\begin{tabular}{c}
$y_1 = 1$ \\
$x_2 = 2$
\end{tabular}}} + -  \downarrow  -+  \underbrace{\downarrow \uparrow}_{\scriptsize{\begin{tabular}{c}
$y_2 = 1$ \\
$x_3 = 1$
\end{tabular}}} + -  \uparrow  -+ \underbrace{\downarrow \downarrow}_{\scriptsize{\begin{tabular}{c}
$y_3 = 2$ \\
$x_1 = 0$
\end{tabular}}}  +-     \rangle. \label{eq:appExampleState}
\end{equation}
It satisfies $(x_1, y_1) = (0, 1)$ as only one of the $\downarrow$ pseudo-spins of the first subsequence of $\uparrow \downarrow \uparrow \downarrow$ has moved the right of the first (leftmost) dipole $-+$, and none to the left of the last (rightmost) dipole $+-$.
As can be straightforwardly observed, it also satisfies $(x_2, y_2) = (2, 1)$ and $(x_3, y_3)=(1, 2)$.

The corresponding Krylov subspace's dimension then reads:
\begin{widetext}
\begin{equation}
\text{dim } \mathcal{K}(\Ket{\Phi_{n}^m})_\mathrm{PBC} = \sum\limits_{x_1, y_1, ...=0}^n \binom{2n-x_1-y_1}{n-y_1} \binom{y_1 + x_2}{y_1} \binom{2n-x_2-y_2}{n-y_2} ... \binom{y_m + x_1}{y_m}.
\end{equation}
\end{widetext}
This can be rewritten as
\begin{equation}
\text{dim } \mathcal{K}(\Ket{\Phi_{n}^m})_\mathrm{PBC} =\sum\limits_{x_1, y_1=0}^n f(x_1, y_1) g_m(y_1, x_1),
\end{equation}
with
\begin{equation}
f(x, y) = \binom{2n-x-y}{n-y}
\end{equation}
and
\begin{align}
g_m(y_1, x_1) &= \sum\limits_{x_2, y_2, ...} \binom{y_1 + x_2}{y_1} \binom{2n-x_2-y_2}{n-y_2} ... \binom{y_m + x_1}{y_m} \notag \\
 &= \sum\limits_{z=0}^n T^{(n)}_{y_1, z} g_{m-1}(z, x_1).
\end{align}
where the $(n+1)\times (n+1)$ transfer matrix $T^{(n)}$  has entries given by
\begin{equation}
T^{(n)}_{x, y} = \sum\limits_{z=0}^n \binom{2n -y - z}{n-z} \binom{x+z}{z} = \binom{2n + 1 + x - y}{n},
\end{equation}
and $x, y = 0, ..., n$.
Defining the matrices $F_{x, y} = f(x, y)$ and $G^0_{x, y} = \binom{x+ y}{x} $, we obtain the simple expression:
\begin{equation}
\text{dim } \mathcal{K}(\Ket{\Phi_{n}^m})_\mathrm{PBC} = \mathrm{Tr}\left( F \left[T^{(n)}\right]^{m-1} G^0 \right).
\end{equation}
Thus, the dimension of the Krylov subspace scales as $(t^{(n)})^{m}$ with $t^{(n)}$ the largest eigenvalue of $T^{(n)}$, as it corresponds to the translation by a single motif $(\uparrow \downarrow)^n - ++ -$.
Using the relation $N = (2n + 4)m$, we readily obtain that the dimension of $\mathcal{K}(\Ket{\Phi_n^m})_\mathrm{PBC}$ 
scales as $(t^{(n)})^{N/(4+2n)}=(t^{(n)})^{L/(8+4n)}$.\\

Let us consider $n=1$ as a concrete example.
The matrix $T^{(1)}$ is given by
\begin{equation}
T^{(1)} = \begin{pmatrix}
3 & 2 \\ 4 & 3
\end{pmatrix}.
\end{equation}
Its eigenvalues are $t^{(1)}_\pm = 3 \pm 2 \sqrt{2}$, and therefore
\begin{align}
\text{dim }  \mathcal{K}(\Ket{\Phi_1^m})_\mathrm{PBC} &\approx (t^{(1)}_+)^m = (t^{(1)}_+)^{\frac{N}{6}} = (t^{(1)}_+)^{\frac{L}{12}}\\
&\approx 1.341^N \approx 1.158^L.
 \end{align} 
The dimension of the Hilbert space cannot be understood from a simple quasi-particle picture.
Indeed, there exists no $p \in \mathbb{N}^*$ such that $(t^{(1)}_+\,^{\frac{1}{12}})^p$ is rational. 
The proof goes as follows: if such a $p$ exists, then there exists $\tilde{p}\in \mathbb{N}^*$ such that $(T^{(1)})^{\tilde{p}}$ has rational eigenvalues.
Eigenvalues of  $(T^{(1)})^{\tilde{p}}$ are roots of the polynomial $\det((T^{(1)})^{\tilde{p}} - \lambda I)$, which has integer coefficients, and a leading coefficient of $1$.
The eigenvalues are therefore real irrational integers, whose intersection with $\mathbb{Q}$ are integers only.
They are given by
\begin{multline}
\frac{1}{2}\left((3-2\sqrt{2})^{\tilde{p}} + (3+2\sqrt{2})^{\tilde{p}} \right)\\
 \pm \sqrt{ \frac{1}{4}\left( (3-2\sqrt{2})^{\tilde{p}} + (3+2\sqrt{2})^{\tilde{p}} \right)^2 - 1 }.
\end{multline}
The first parenthesis is trivially an integer, while the second term is of the form $\sqrt{A^2-1}$, with $A\in \mathbb{N}$ and $A>1$.
This second term is therefore never an integer, and $(T^{(1)})^{\tilde{p}}$ can have neither integer nor rational eigenvalues.\\

For open boundary conditions, the Krylov subspace dimension can be obtained from a similar formula:
\begin{widetext}
\begin{align}
\text{dim }  \mathcal{K}(\Ket{\Phi_n^m})_\mathrm{OBC}&= \sum\limits_{y_1, ...} \binom{2n-y_1}{n} \binom{y_1 + x_2}{y_1} \binom{2n-x_2-y_2}{n-y_2} ... \binom{2n-x_m}{n} \\
&= \sum\limits_{y_1, x_m} f(0, y_1) g_m(y_1, x_m) f(x_m, 0),
\end{align}
\end{widetext}
The dimension of the OBC Krylov subspace therefore scales as in the periodic case.\\

In Tables~\ref{tab:HilbertSpaceDim-Sequences1} and \ref{tab:HilbertSpaceDim-Sequences2}, we summarize the dimensions of the Krylov subspace and of the reduced density matrix for a half-chain cut.

\begin{table*}
\begin{tabular}{|c|c|c|c|c|c|c|c|c|}
 \hline m  & $N^\mathrm{PBC}$ & $\dim \mathcal{H}^\mathrm{PBC}$ & $\mathcal{C}(\mathcal{H}^\mathrm{PBC})$ & $\dim \rho_{\frac{1}{2}}^\mathrm{PBC}$ & $N^\mathrm{OBC}$ & $\dim \mathcal{H}^\mathrm{OBC}$ & $\mathcal{C}(\mathcal{H}^\mathrm{OBC})$ & $\dim \rho_{\frac{1}{2}}^\mathrm{OBC}$\\
\hline 1 & 6 & 6 & 2.167 & 5 & 8 & 10 & 2.7 & 4\\
\hline 2 & 12 & 34 & 4.03 & 12 & 14 & 58 & 4.60 & 10 \\
\hline 3 & 18 & 198 & 6.01 & 29 & 20 & 338 & 6.59 & 24 \\
\hline 4 & 24 & 1154 & 8 & 70 & 26 & 1970 & 8.59 & 58\\
\hline 5 & 30 & 6726 & 10 & 169 & 32 & 11482 & 10.59 & 140 \\
\hline 6 & 36 & 39202 & 12 & 408 & 38 & 66922 & 12.59 & 338\\
\hline
\end{tabular}
\caption{We summarize the properties of the Krylov space defined for the generating sequence $(\uparrow \downarrow - + + -)^m$. The column $3$ to $5$ ($6$ to $8$ are for the OBC (PBC) case. Note that for OBC, we add a sequence $\uparrow \downarrow$ at the right end of the generating state for symmetry. The third and sixth columns list the dimensions of the Krylov subspaces. The dimension of the Krylov spaces approximately grows as $1.341^N$. The fourth and seventh columns show the connectivity of the Hamiltonian $\mathcal{C}$ (here defined as the ratio of the number of non-zero non-diagonal terms in the Hamiltonian over the Hilbert space dimension). Finally, the fifth and eighth columns list the dimensions  $\dim \rho_{\frac{1}{2}}^\mathrm{OBC}$ of the reduced density matrix for a cut exactly in the middle of the chain.}
\label{tab:HilbertSpaceDim-Sequences1}
\end{table*}

\begin{table*}
\begin{tabular}{|c|c|c|c|c|c|c|c|c|}
 \hline m  & $N^\mathrm{PBC}$ & $\dim \mathcal{H}^\mathrm{PBC}$ & $\mathcal{C}(\mathcal{H}^\mathrm{PBC})$ & $\dim \rho_{\frac{1}{2}}^\mathrm{PBC}$ & $N^\mathrm{OBC}$ & $\dim \mathcal{H}^\mathrm{OBC}$ & $\mathcal{C}(\mathcal{H}^\mathrm{OBC})$ & $\dim \rho_{\frac{1}{2}}^\mathrm{OBC}$\\
\hline 1 & 8 & 30 & 3.23 & 19 & 12 & 126 & 4.72 & 16\\
\hline 2 & 16 & 786 & 6.32 & 96 & 20 & 3441 & 7.81 & 91 \\
\hline 3 & 24 & 21873 & 9.47 & 514 & 28 & 96054 & 10.97 & 472 \\
\hline
\end{tabular}
\caption{We summarize the properties of the Krylov space defined for the generating sequence $(\uparrow \downarrow \uparrow \downarrow - + + -)^m$. The column $3$ to $5$ ($6$ to $8$ are for the PBC (OBC) case. Note that for OBC, we add a sequence $\uparrow \downarrow$ at the right end of the generating state for symmetry. The third and sixth columns list the dimensions of the Krylov subspaces. The dimension of the Krylov spaces approximately grows as $1.516^N$. The fourth and seventh columns show the connectivity of the Hamiltonian $\mathcal{C}$ (here defined as the ratio of the number of non-zero non-diagonal terms in the Hamiltonian over the Hilbert space dimension). Finally, the fifth and eighth columns list the dimensions  $\dim \rho_{\frac{1}{2}}^\mathrm{OBC}$ of the reduced density matrix for a cut exactly in the middle of the chain.}
\label{tab:HilbertSpaceDim-Sequences2}
\end{table*}

\section{Scaling of the reduced density matrix and entanglement properties of $\Ket{\Phi^m_1}$} \label{app:redmat}

In the Krylov subspace built from $\Ket{\Phi^m_1}$, the reduced density matrix follows an interesting simple pattern.
As can be seen in Fig.~\ref{fig:entanglementGrowth}\href{fig:entanglementGrowth}{a}, each cut $l=3j+1$, $l=3j+2$ and $l=3j+3$ has the same entropy for $j\geq 1$ and PBC, i.e.,
\begin{equation}
S(l_\Aa = 3j+1) = S(l_\Aa = 3j+2) = S(l_\Aa = 3j+3). \label{eq:appPlateauEntr}
\end{equation}
For OBC, this property is true already at $j=0$.
In this Appendix, we show that the reduced density matrices obtained for these cuts are actually identical.
It can be proven by rewriting $\Ket{\Phi^m_1}$ in terms of states combining three consecutive spins.
We use the compact notation $\Ket{\nu_1 \nu_2 \nu_3} = \Ket{4\nu_1 + 2\nu_2 + \nu_3}$ where $\nu_j = 0$ or $1$.
In this notation, $\Ket{\Phi^m_1}$ can be written as $\Ket{(4474)^m}$.
The relevant transformation rules induced by the pair-hopping terms now read 
\begin{equation}
\Ket{44} \leftrightarrow \Ket{30} \text{ and } \Ket{47} \leftrightarrow \Ket{33}\label{eqapp:transforules}
\end{equation}
All other configurations of $\Ket{0}$, $\Ket{3}$, $\Ket{4}$ and $\Ket{7}$ are cancelled by the four fermions hopping terms and preserved by the transverse field.
No state containing $\Ket{1}$, $\Ket{2}$, $\Ket{5}$, $\Ket{6}$ and $\Ket{8}$ is therefore connected to $\Ket{\Phi^m_1}$.
The Krylov subspace only contains (a subset of the) states built from the triplets $\Ket{0}$, $\Ket{3}$, $\Ket{4}$ and $\Ket{7}$.\\

For clarity, we start with the OBC case, and discuss the PBC case later.
We will consider separately the case $j=0$, $j=1$ and $j=2$ by direct inspection.
Then we will address the generic $j$ value.
For any given starting state, we can write the single-site density matrix of the left most state explicitely as:
\begin{equation}
\rho(l_\mathcal{A}=1) = \alpha \Ket{0}\Bra{0} + \beta \Ket{1}\Bra{1} + (\gamma \Ket{0}\Bra{1} + h.c.),\label{eq:defineLeftRho}
\end{equation}
where $\alpha, \beta$ are positive real coefficients and $\gamma$ is a complex constant.
Using the transformation rules in Eq.~\eqref{eqapp:transforules}, the configurations $\Ket{44}$ can only be mapped to $\Ket{30}$ or $\Ket{43}$.
This means the configuration of the first triplet is either $\Ket{3}$ or $\Ket{4}$.
Once we fix the first spin, the next two are determined.
Using Eq.~\eqref{eq:defineLeftRho}, we obtain the following expressions for the reduced density matrices at $l_\Aa=2$ and $l_\Aa=3$.
\begin{equation}
\rho(l_\mathcal{A}=2) = \alpha \Ket{01}\Bra{01} + \beta \Ket{10}\Bra{10} + (\gamma \Ket{01}\Bra{10} + h.c.),\\
\end{equation}
\begin{multline}
\rho(l_\mathcal{A}=3) = \alpha \Ket{011}\Bra{011} + \beta \Ket{100}\Bra{100}\\
 + (\gamma \Ket{011}\Bra{100} + h.c.).
\end{multline}
The coefficients of the density matrices, and therefore the vNEE, remain the same whether we cut after the first, second or third spin.
Now, we turn to a cut through the second triplet (corresponding to $j=1$ in Eq.~\eqref{eq:appPlateauEntr}).
The triplet itself can be either $\Ket{4}$, $\Ket{3}$ or $\Ket{0}$.
On the other hand, also taking into account the first triplet, lead to the following three combinations
\begin{equation}
\Ket{44} = \Ket{1001\,00},\ \Ket{43} = \Ket{1000\,11}\text{ and }  \Ket{30} = \Ket{0110\,00}.
\end{equation}
Therefore, fixing the first $4$ spins, i.e., the first triplet and the first spin of the second triplet again entirely determines the states of the second and third spin.
Eq.~\eqref{eq:appPlateauEntr} is therefore also valid for $j=1$.
A similar reasoning can be applied to the third triplet and $j=2$, with the sequences:
\begin{equation}
\Ket{307},\ \Ket{433},\ \Ket{434} \text{ and } \Ket{447}.
\end{equation}
To straightforwardly extend the results to the rest of the system, it is enough to consider all possible four triplets sequences in the Krylov subspace.
There are only $44$ such sequences (out of $2^{12} = 4096$ possible spin configuration and $4^4 = 256$ combination of triplets), given in Tab.~\ref{tab:listTriplets}.
They can be obtained by brute force for $m=3$ ($m=2$ for PBC), and the limited propagation of the dipoles through the pseudo-spins ensures that no other configurations arise for larger systems.
For all these sequences, fixing the first three triplets and the first spin of the fourth triplet is enough to determine the whole sequence.
Eq.~\eqref{eq:appPlateauEntr} is therefore valid for any $j$.
For PBC, the same analysis can be performed, leading to the same property and pattern observed in Fig.~\ref{fig:entanglementGrowth}\href{fig:entanglementGrowth}{a}.
The only difference is that the reduced density matrix does change going from the first spin to the second spin as the left-most triplet can be $\Ket{4}$, $\Ket{3}$ and $\Ket{0}$.
For further spins and triplets, the proof is similar to the one we derived for OBC.\\

\begin{table}
\begin{tabular}{|c|c|c|c|c|c|c|c|}
\hline $0 3 3 3$ & $0 7 4 4$ & $3 3 0 4$ & $3 4 7 0$ & $4 4 4 7$ & $4 7 4 4$ & $7 4 4 3$\\
$0 3 3 4$ & $3 0 3 3$ & $3 3 3 0$ & $4 3 0 7$ & $4 4 7 3$ & $7 0 3 3$ & $7 4 4 4$\\
$0 3 4 7$ & $3 0 3 4$ & $3 3 4 3$ & $4 3 3 3$ & $4 4 7 4$ & $7 0 3 4$ &\\
$0 4 7 3$ & $3 0 4 7$ & $3 3 4 4$ & $4 3 3 4$ & $4 7 0 3$ & $7 0 4 7$ &\\
$0 4 7 4$ & $3 0 7 3$ & $3 4 3 0$ & $4 3 4 7$ & $4 7 0 4$ & $7 3 0 3$ &\\
$0 7 3 0$ & $3 0 7 4$ & $3 4 4 3$ & $4 4 3 3$ & $4 7 3 0$ & $7 3 0 4$ &\\
$0 7 4 3$ & $3 3 0 3$ & $3 4 4 4$ & $4 4 3 4$ & $4 7 4 3$ & $7 4 3 0$ &\\ \hline
\end{tabular}
\caption{List of the $44$ possible combinations of triplets of spins arising in $\Ket{\Phi^m_1}$ }
\label{tab:listTriplets}
\end{table}

Additionally, the growth of the reduced Hilbert space from one three-site plateau to the next is very irregular, as shown in Fig.~\ref{fig:entanglementGrowth}\href{fig:entanglementGrowth}{a}.
The alternating small and large jumps in entropy observed in the ergodic phase translates into alternating low and high entanglement plateaus in the MBL phase at strong disorder.
Both phenomena can be explained by simple perturbative expansion arguments in the triplet language.
At strong disorder, the dominant energy terms are the disorder terms, and we can assume that the eigenstates are generally close to product states in the $z$ spin-basis, and therefore in the triplet basis.
Non-zero contributions to the entropy mainly come from local resonances, with two nearest neighbours triplet forming a pair due to the corresponding hopping term.
In Table~\ref{tab:Pair_ud-++-}, we summarize how many pairs of states the hopping terms can generate depending on where they are applied on the Krylov spaces generated by $\Ket{\Phi_1^m}$.\\

We treat as an example the case $m=1$.
With periodic boundary conditions, the Krylov subspace consists of only $6$ states: $\ket{i} \equiv \Ket{4474}$, $\Ket{ii} \equiv \Ket{3074}$, $\Ket{iii} \equiv \Ket{4334}$, $\Ket{iv} \equiv \Ket{0473}$, $\Ket{v} \equiv \Ket{0333}$ and $\Ket{vi} \equiv \Ket{0347}$.
The pair-hopping terms linking the first two triplets only transform the state $\ket{i}$ into $\Ket{ii}$ (and vice versa).
Similarly, those connecting the $3\mathrm{rd}$ and $4\mathrm{th}$ triplets only transform $\Ket{v}$ into $\Ket{vi}$.
On the other hand, the pair-hopping terms connecting the $2\mathrm{nd}$ and the $3\mathrm{rd}$ triplet map $\Ket{i}$ into $\Ket{iii}$ and $\Ket{iv}$ into $\Ket{vi}$.
The alternating low and high number of pairs perfectly explain the observed entropy patterns.
If this number is small, the average entropy in the MBL phase is lower as a limited number of local resonant pairs can exist.
Conversely, at low disorder, the number of pairs reflect the number of connections in the configuration basis, that is to say the growth of the reduced Hilbert space  when increasing system size.
A lower number of pairs implies that the subspace grows less and therefore that the entropy increases less in the ergodic phase.


\begin{table}
\begin{tabular}{|c|c|c|c|c|}
\hline Hopping term $j \leftrightarrow j+1$& $0-1$ & $1-2$ & $2-3$ & $3-4$\\
\hline Number of pairs of states in $\Ket{\Phi_1^1}$& $1$ & $2$ & $1$ & $2$ \\
\hline Number of pairs of states in $\Ket{\Phi_1^2}$& $5$ & $12$ & $5$ & $12$  \\
\hline Number of pairs of states in $\Ket{\Phi_1^3}$& $29$ & $70$ & $29$ & $70$  \\
\hline Number of pairs of states in $\Ket{\Phi_1^4}$& $169$ & $408$ & $169$ & $408$  \\
\hline Number of pairs of states in $\Ket{\Phi_1^5}$ & $985$ & $2378$ & $985$ & $2378$  \\
\hline
\end{tabular}
\caption{Number of pairs of states created by the hopping terms connecting the triplet $j$ (that is to say the sites $3j+1$, $3j+2$ and $3j+3$)  and the triplet $j+1$ for the Krylov spaces generated from the state $\Ket{\Phi^m_1}$ for different values of $m$, taking PBC.
The remaining terms can be obtained due to the invariance by translation of $12$ sites.
Due to the pattern of the generating state, we alternate between large and small numbers of connected states. 
This pattern explains the alternating large and small increase of entropies in the ergodic phase, and the alternating high and low entropy links in the MBL phase seen in Fig.~\ref{fig:entanglementGrowth}\href{fig:entanglementGrowth}{a}.}
\label{tab:Pair_ud-++-}
\end{table}

\section{Additional numerical data}\label{app:AdditionalData}
In this Appendix, we present briefly some additional numerical results mentioned in the main text.

\subsection{Mean level spacing}\label{app:AdditionalData_meanlevel}
\begin{figure}[ht!]
\includegraphics[width=0.9\linewidth]{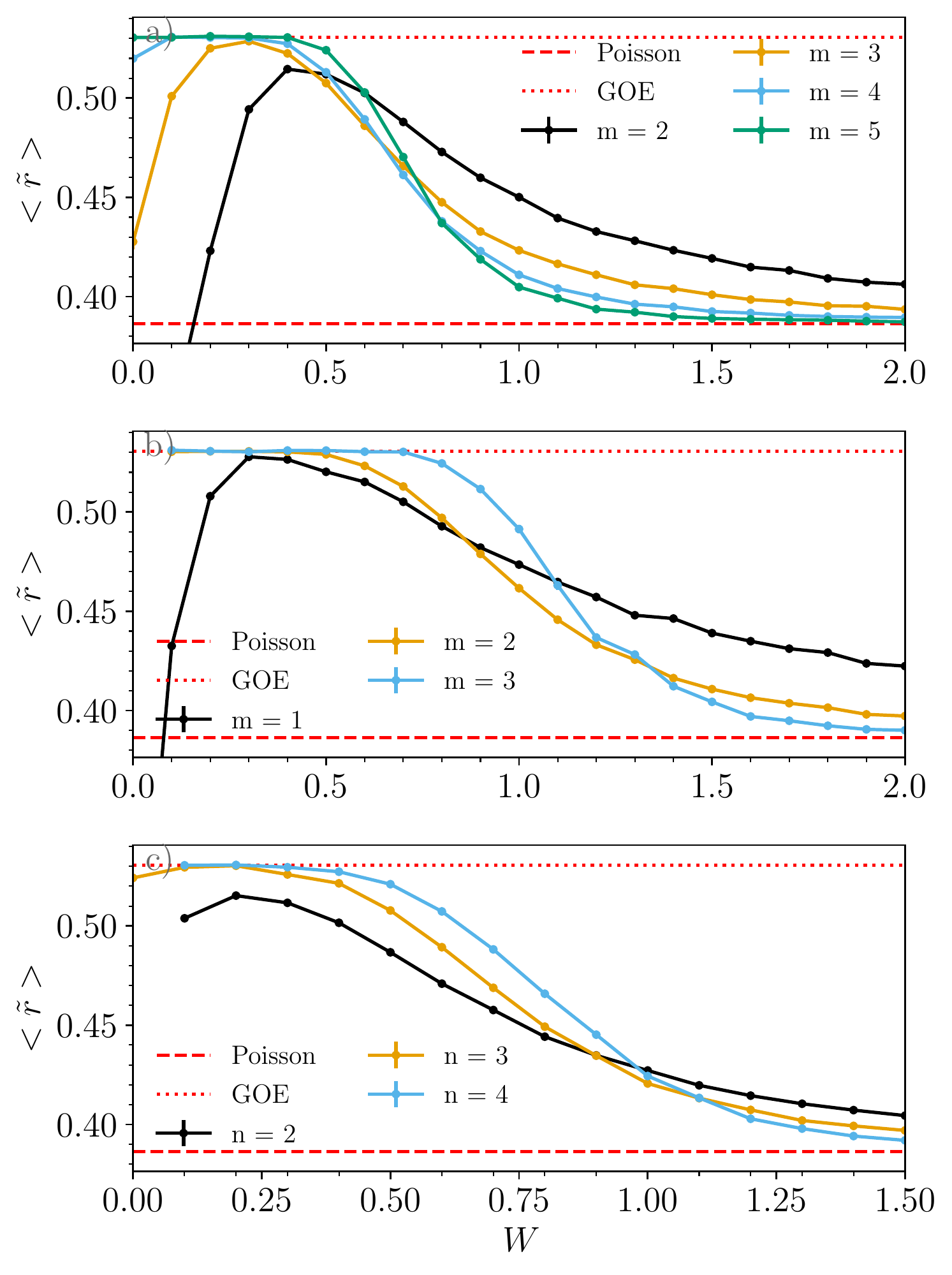}
\caption{Mean value of the level spacing ratio for the three families $\Ket{\Phi^m_1}$, $\Ket{\Phi^m_2}$ and $\Ket{\Psi^1_n}$ for different values of $m$ and $n$. We generically observe a crossover from the GOE value towards the Poisson value. The observed behavior is compatible with the results based on the KL-divergence in the main text.
}
\label{fig:MLRatio}
\end{figure}

We first turn towards the computation of the mean value of the energy level ratio.
The mean level ratio is simply defined as the average of the distribution introduced in Eq.~\eqref{eq:deftilder}.
It is a good indicator of the MBL phase transition as it crosses from $0.5307$ for a GOE distribution to $0.38629$ for Poisson distribution\citep{Oganesyan2007}.
On the other hand, it generally shows a significant shift with system size and captures only partially the behavior of the distributions through the phase transition.
In Fig.~\ref{fig:MLRatio}, we represent the mean level ratio for the three Krylov subspaces we consider.
We observe in all cases, a crossover from GOE statistics to Poisson statistics.
The behavior in each family is qualitatively different, with a sharper transition for $\Ket{\Phi^m_1}$, significant changes with system size for $\Ket{\Phi^m_2}$ (compared to the other two Krylov subspaces), and a slower transition for $\Ket{\Psi_n^1}$.
We also observe a significant shift  of the transition point with system size for the family generated by $\Ket{\Psi_n^1}$.
These results are consistent with those obtained by studying the KL-divergence in the main text.

\subsection{Mobility edge}\label{app:AdditionalData_mobedge}
\begin{figure}[ht!]
\includegraphics[width=0.9\linewidth]{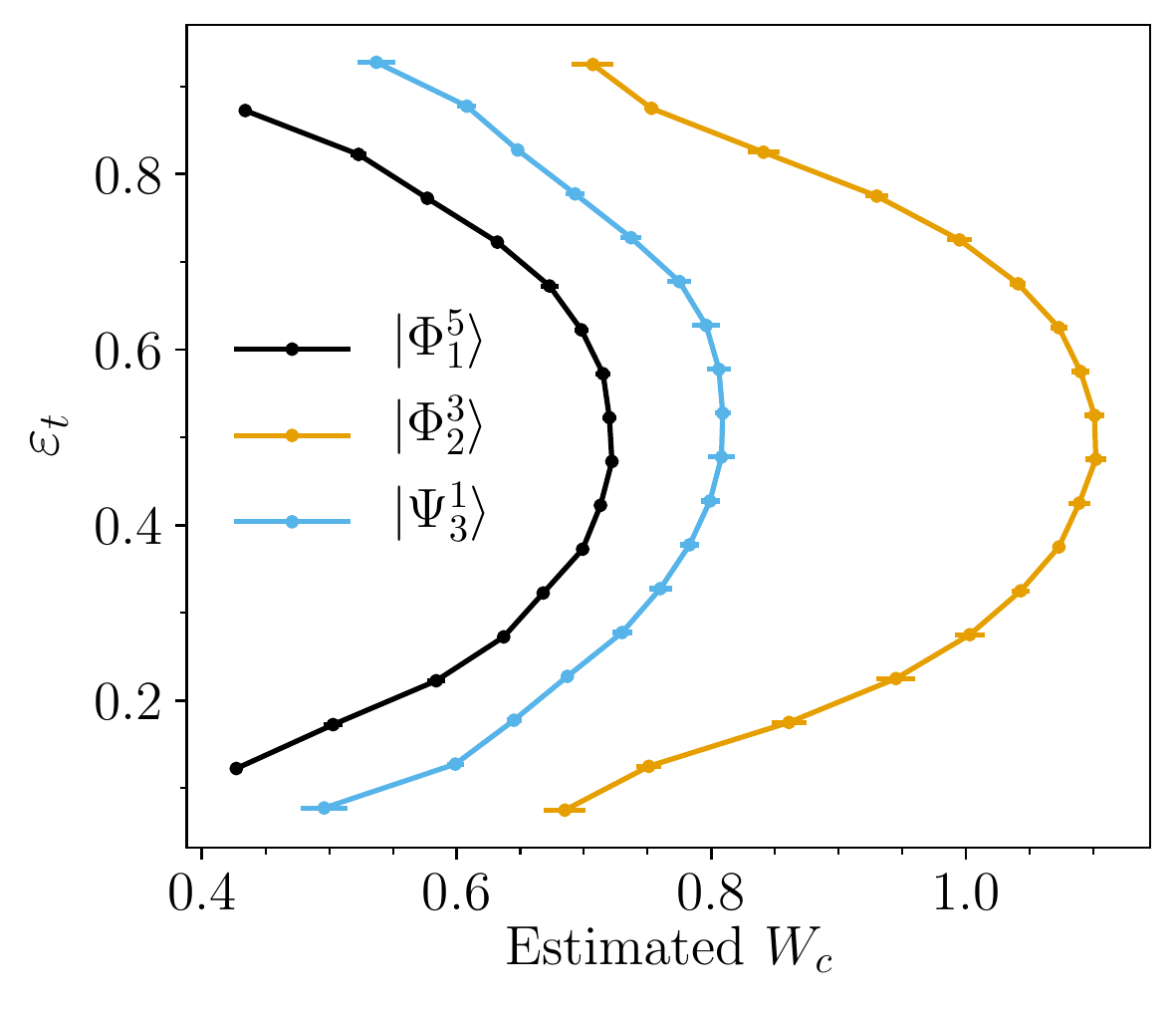}
\caption{Estimated critical disorder strengths $W_c$ for the three families $\Ket{\Phi^m_1}$ (a), $\Ket{\Phi^m_2}$ (b) and $\Ket{\Psi^1_n}$ (c) for different values of $m$ and $n$, as a function of the normalized energy. We compute the distribution of level ratio of states with normalized energies $\varepsilon \in [\varepsilon_t - 0.025, \varepsilon_t + 0.025]$ for a range of $\varepsilon_t$. We estimate the critical disorder strength using the KL-divergence as shown in Sec.~\ref{sec:LS-singleK}. For all three families, we observe a significant variation of the transition point with energy level. Error bars are obtained through subsampling of our data.}
\label{fig:MobilityEdge}
\end{figure}

As was observed\citep{Gornyi2005, Basko2006, Kjall2014, Luitz2015} in similar models, a MBL transition typically presents a mobility edge, that is to say that the critical disorder strength depends on the energy of the eigenstates.
The same mobility edge can also be observed in our constrained model, as illustrated in Fig.~\ref{fig:MobilityEdge}.
We compute the level spacing ratio distribution of states of normalized energies $\varepsilon \in [\varepsilon_t - 0.025, \varepsilon_t + 0.025]$ for $\varepsilon_t$ varying from $0.075$ to $0.925$.
We then estimate the critical disorder strengths as shown in Sec.~\ref{sec:LS-singleK}: it corresponds to the disorder strengths where the KL-divergences of the numerical distribution with the Poisson distribution or GOE one are equal.
For all three spaces, we observe that the critical disorder strength significantly varies with the energy target, over a range and shape comparable to those observed in other models\citep{Luitz2015}.

\subsection{Level spacing ratio for two mixed Krylov subspaces}\label{app:AdditionalData_bothStats}
\begin{figure}
\subfloat{\includegraphics[width=0.9\linewidth]{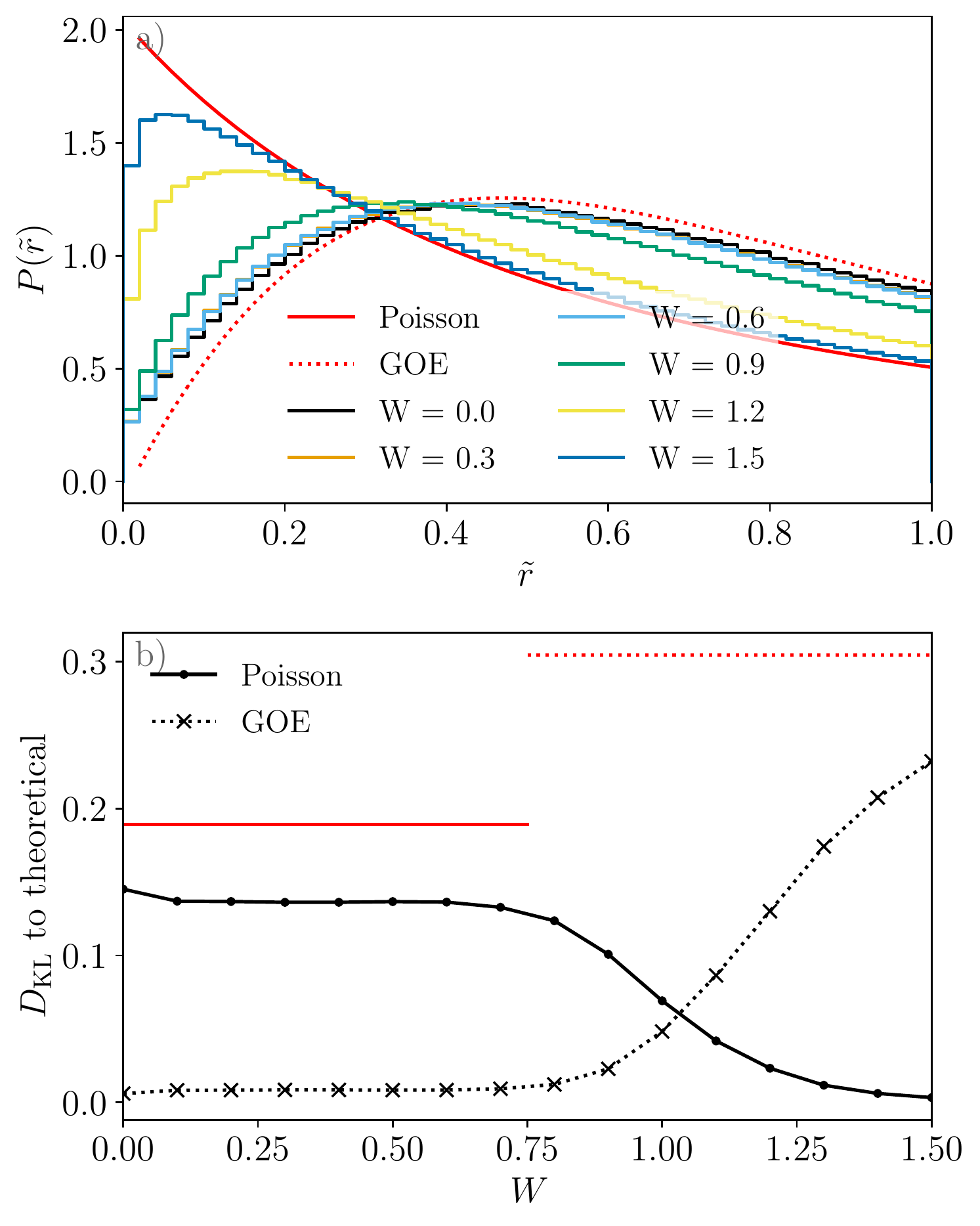}}
\caption{a) Distribution of the level spacing ratio when mixing the Krylov subspaces generated by $\Ket{\Phi_1^4}$ and $\Ket{\Phi_2^3}$, corresponding to $N=24$. We observe a significant divergence from the ergodic GOE distribution at zero disoder, which converges slowly towards the Poisson distribution at higher disorder. b) KL-divergence of that distribution with the two reference distributions. The red lines mark the divergences between $P_\mathrm{num}$ and $P_\mathrm{Poi})$. The discrepancy at low disorder is better seen in $D_\mathrm{KL}(P_\mathrm{num}, P_\mathrm{Poi})$ than in $D_\mathrm{KL}(P_\mathrm{num}, P_\mathrm{GOE})$. The phase transition is not as sharp than in Sec.~\ref{sec:LS-singleK} (see Fig.~\ref{fig:LS}). Error bars are too small to be seen.
}
\label{fig:bothStatistics}
\end{figure}

Finally we study the level spacing ratio statistics obtained from mixing two Krylov subspaces seeing a transition at different disorder strengths.
More precisely here, we consider the two Krylov subspaces generated by $\Ket{\Phi_1^4}$ (of dimension $1154$) and $\Ket{\Phi_2^3}$ (of dimension $21873$), taken with periodic boundary conditions with $N=24$.
We compute the level spacing ratio distributions by mixing the spectra obtained in the two spaces for the same disorder realisations.
In Fig.~\ref{fig:bothStatistics}, we show our results for eigenstates with $\varepsilon \in [0.4, 0.6]$, averaging over $200$ disorder realizations.
The distribution of the level spacing ratio shows a significant departure from the GOE distribution at low disorder and converges towards the Poisson distribution at higher disorder.
The behavior of the KL-divergence between the numerical distributions and our two reference distributions also shows a qualitative and quantitative difference with the distributions studied in Sec.~\ref{sec:LS-singleK} (see Fig.~\ref{fig:LS}\href{fig:LS}{a-b}).
The difference at low disorder is better seen when studying $D_\mathrm{KL}(P_\mathrm{num}, P_\mathrm{Poi})$, with a saturation value significantly lower than $D_\mathrm{KL}(P_\mathrm{GOE}, P_\mathrm{Poi})$.
We also observe a cross-over to the Poisson distribution, with a critical disorder strength in between the ones obtained in both subspaces.
Unsurprisingly, the transition also appears much less sharp than in our study of the isolated Krylov spaces.

\bibliography{Disorder}

\end{document}